\documentclass[10pt,twocolumn,letterpaper]{article}

\usepackage{cvpr}              % To produce the CAMERA-READY version

\makeatletter
\@namedef{ver@everyshi.sty}{}
\makeatother
\usepackage{pgfplots}
\pgfplotsset{compat=1.16}

\usepackage{graphicx}
\usepackage{amsmath}
\usepackage{amssymb}
\usepackage{booktabs}
\usepackage{multirow}
\usepackage{cuted}
\usepackage[many]{tcolorbox}
\newtcolorbox{mybox}[1][]{
  boxsep=0pt,
  left=-10pt,right=2pt,top=2pt,bottom=2pt,
	enhanced,
	colback=blueish!5!white,
	colframe=blueish,
	#1
}

\makeatletter
\def\blfootnote{\gdef\@thefnmark{}\@footnotetext}
\makeatother

\usepackage{fixltx2e}
\usepackage{musicography}
\usepackage[accsupp]{axessibility}

\usepackage[pagebackref,breaklinks,colorlinks]{hyperref}

\usepackage[capitalize]{cleveref}
\crefname{section}{Sec.}{Secs.}
\Crefname{section}{Section}{Sections}
\Crefname{table}{Table}{Tables}
\crefname{table}{Tab.}{Tabs.}

\everypar{\looseness=-1}
\def\cvprPaperID{11001} % *** Enter the CVPR Paper ID here
\def\confName{CVPR}
\def\confYear{2022}

\usepackage{overpic}
\usepackage{enumitem} %< control spacing in itemize/enumerate/...
\usepackage{overpic} %< add raw math symbols to figures
\usepackage{color}

\usepackage{tikz}
\usetikzlibrary{tikzmark}
\usepackage[tikz]{bclogo}
\usepackage{array}
\usepackage{soul}
\newcommand{\videocorpus}{\mathcal{V}}

\newcommand{\video}{x^v}
\newcommand{\music}{x^m}
\newcommand{\numsegments}{L}
\newcommand{\indexsegment}{l}
\newcommand{\targetsize}{N}
\newcommand{\outputvideo}{y^v}
\newcommand{\outputmusic}{y^m}

\definecolor{turquoise}{cmyk}{0.65,0,0.1,0.3}
\definecolor{purple}{rgb}{0.65,0,0.65}
\definecolor{dark_green}{rgb}{0, 0.5, 0}
\definecolor{orange}{rgb}{0.8, 0.6, 0.2}
\definecolor{red}{rgb}{0.8, 0.2, 0.2}
\definecolor{darkred}{rgb}{0.6, 0.1, 0.05}
\definecolor{blueish}{rgb}{0.0, 0.3, .6}
\definecolor{light_gray}{rgb}{0.7, 0.7, .7}
\definecolor{pink}{rgb}{1, 0, 1}
\definecolor{greyblue}{rgb}{0.25, 0.25, 1}

\usepackage{blindtext}

\renewcommand{\paragraph}[1]{\vspace{1em}\noindent\textbf{#1}.}
\begin{document}
% \title{Self-supervised Audio-Visual Learning for Recommending Music for Video}
% \title{On the Importance of Temporal Context for Corresponding Video and Music}
\title{It's Time for Artistic Correspondence in Music and Video}

\author{D\'idac Sur\'is\textsuperscript{\musEighth} \\ %\eighthnote}\\
Columbia University\\
{\tt\small didacsuris@cs.columbia.edu}
\and
Carl Vondrick\\
Columbia University\\
\and
Bryan Russell\\
Adobe Research\\
\and
Justin Salamon\\
Adobe Research\\
}
% See cvpr.sty to change the format of this
\date{\href{http://musicforvideo.cs.columbia.edu}{\textbf{\textcolor{purple}{\texttt{musicforvideo.cs.columbia.edu}}}}}

%%% brussell: uncomment this next "maketitle" line and comment the "twocolumn" section below if you want to remove the two-column teaser on page 1.

% \maketitle

\twocolumn[{
\renewcommand\twocolumn[1][]{#1}
\maketitle
% \vspace{-1.9cm}
\vspace{-1cm}
\begin{center}
\includegraphics[width=1\textwidth]{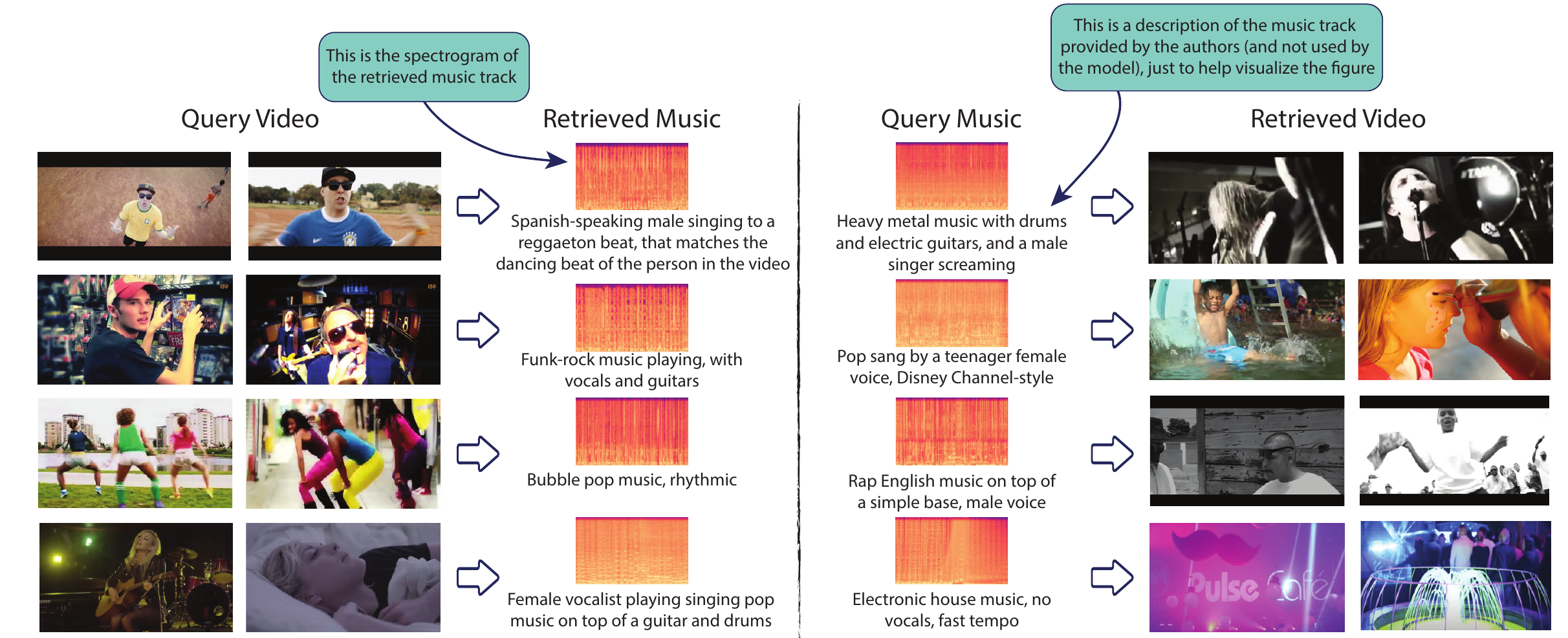}
\vspace{-0.4cm}
\captionof{figure}{
We present an approach for recommending a music track for a given video, and vice versa.  We model the long-term temporal context of both signals, allowing our model to capture the high-level artistic correspondences between them.  Our model learns a strong audiovisual representation that allows us to retrieve videos and music that look and sound natural to humans. On the left we show query video segments with the corresponding retrieved music segments, and on the right we show the opposite retrieval direction. Our model's audiovisual correspondence exploits artistic attributes such as music genre or rhythm. %See \href{http://musicvideo.cs.columbia.edu}{our website} for playable visualizations.
}
\label{fig:musicvideo_retrieval}
\end{center}%
}]

\begin{abstract}
\blfootnote{\textsuperscript{\musEighth}Work partly done during an internship at Adobe Research.}
   We present an approach for recommending a music track for a given video, and vice versa, based on both their temporal alignment and their correspondence at an artistic level. 
   We propose a self-supervised approach that learns this correspondence directly from data, without any need of human annotations.
   In order to capture the high-level concepts that are required to solve the task, we propose modeling the long-term temporal context of both the video and the music signals, using Transformer networks for each modality. Experiments show that this approach strongly outperforms alternatives that do not exploit the temporal context. The combination of our contributions improve retrieval accuracy up to 10$\times$ over prior state of the art. This strong improvement allows us to introduce a wide range of analyses and applications. For instance, we can condition music retrieval based on visually defined attributes. %.\ds{something like "such as music + conditioning images equals new music"}.
%   \JS{depending on the number of applications and how impactful we ultimately feel they are we may want to list them here. Alternatively if we feel they are weak we may to end on something else e.g.~bias analysis in the context of the underexplored problem of artistic audiovisual correspondence learning.}
   % Finally, we also make an extensive analysis to study what signals are important for music-video correspondence, from low-level ones (like tempo or brightness), to high-level ones (like music genre).
   %Finally, due to its symmetry, our model is also capable of performing the task in the opposite direction. Specifically, given a set of video clips and a long music track, the model is capable of rearranging the video clips so that they align with the audio.
   
%   \begin{mybox}
% \color{blueish}\bcattention\begin{minipage}{0.9\columnwidth}\small{\bf{We strongly encourage the reader to view the supplementary material where our video+music results can be fully appreciated.}}\end{minipage}
% \end{mybox}

\end{abstract}
% \vspace{-1.2cm}
\section{Introduction}
\vspace{-0.2cm}
\label{sec:intro}

% (1) Start by stating which problem you are addressing, keeping the audience in mind. They must care about it, which means that sometimes you must tell them why they should care about the problem.
%
% - Potential first-liner (needs a lot more iteration - but goal is to lead off with the task + A/V attribute and genre question): When video content creators search for relevant music for a given video, there are often very subtle decisions being made. For example, what kinds of music genres pair well with the audiovisual attributes (e.g., describe examples here) depicted in the video?

Music is a crucial component of video creation, for example soundtracks in feature film, music for advertisements, background music in video blogs, or creative uses of music in social media. However, choosing the right music for a video is difficult---the video creator needs to determine what kind of music to use for different moments in the video and then search for this music. Each of these tasks presents difficulties: choosing the right music to set the mood of a video can be hard for non-professionals and, even when you know what type of music you want, it can be hard to search for it using conventional text-based methods. It is very hard to describe the ``feel'' of a song in words and metadata-based search engines are not well suited for this task.  

An automated tool to suggest relevant music given video footage as input could be of great value to a range of creators, from beginners and amateurs in need of a simple solution, through to communicators and professionals in search of inspiration. 
The inverse problem, matching video footage to a given song, similarly presents notable challenges, and a solution has the potential to unlock new creative applications. As such, an automated tool to perform retrieval in both directions, from video to music and vice versa, is of great interest.

While other audio-visual tasks aim to establish {\em physical correspondences} for discrete events between the two modalities (\eg, the sound of a person clapping with the visual motion of the person performing the clapping action)~\cite{Arandjelovic2017a,alayrac_self-supervised_2020,Alwassel2019}, such correspondences are predominantly not the deciding factor for pairing music with video. 
The determining factors for the pairing task are instead often ``artistic'' and non-physical, and may comprise the overall visual style or aesthetics of the video, and the genre, mood or ``feel'' of the music. 
Additionally, a system may pair musical genres with visual attributes (\eg, depicted scene type or musical instrument played) or populations presenting a particular gender or race. 
Studying the interplay of these factors is important for understanding and exposing how a system makes its decisions and mitigating potential bias~\cite{paper_about_gender_imbalance_music_recommenders}. 

To address these tasks, we seek to train an audio-visual model to determine how well a paired video and music audio clip ``go together'', or correspond, where we learn this correspondence directly from video data without requiring any manual labeling. Once trained, the model can be used to retrieve music that would pair well with a given input video, and to retrieve video clips that would pair well with a given input music track (see Fig.~\ref{fig:musicvideo_retrieval} for some examples). 
Moreover, we seek to understand how a trained model associates the aforementioned musical genres and visual attributes. 

As it is difficult to manually collect annotated data at large scale describing the mood of video and musical audio, we leverage self-supervision, \ie, learning from the structure inherent to the data. Since we have access to large video collections where music and video have already been paired together by human creators, we %can 
leverage these data to learn what makes for a good pairing. The model is presented with both the large collection paired by human creators and randomly paired audio/video tracks, and is trained to distinguish between the two collections.

Previous approaches for this task typically rely on corresponding short video and music segments or aggregating features over multiple segments~\cite{pretet_cross-modal_2021}.
However, as the correspondence between video and music is often an artistic one, it often depends on long-range temporal context, which is hard to capture in a short segment or by aggregating multiple segment features. 
For example, a given scene in a movie conditions the ``mood'' of the music in the next scene, and the proximity of the climax of a song conditions how the video clip is edited \cite{pretet2021there,Li2019QueryBV,CBVMR}.

Furthermore, these prior approaches optimize metric losses that do not weight hard examples during training~\cite{chen2020simple}, and leverage modality-specific visual base features trained on a fixed-vocabulary classification task~\cite{deng2009imagenet} or audio base features that are not specific for music\cite{openl3}. 

Finally, while these approaches evaluate retrieval accuracy, they do not study how a model associates musical genre and visual attributes.

To address these challenges, we make the following contributions. 
First, we show for the first time that temporal context is important for this artistic correspondence learning task. We do so by leveraging a Transformer architecture~\cite{Vaswani2017} to model long-range temporal context and employing other best practices for video-music retrieval (\eg, optimize a contrastive loss during training, build on strong base features for each modality), leading to a dramatic $10\times$ improvement in retrieval accuracy. 
Second, we conduct a detailed analysis of our model, shedding light
onto what visual attributes present in the video, such as scene type and musical instruments, are used by the model to establish artistic correspondence with different musical genres. This analysis includes ``attributes'' whose over-simplistic definition or representation such as gender and race can lead to potentially concerning biases.
Third, we demonstrate the usefulness of the learned audiovisual representation through several applications, including novel ones such as combining a music query with visual attributes to retrieve music of the same genre where the visual attributes are musically represented in the audio signal. Finally, we study and discuss potential issues with our model related to bias. Since our task is concerned with learning artistic correspondence based on video-music pairings made by humans, rather than audio-visual correspondence grounded in physics, it presents new and important challenges and considerations concerning bias, cultural awareness and appropriation.

\section{Related Work}
\label{sec:related}

\textbf{Music from video.}
Several frameworks have been proposed to recommend music for a given video. However, most of them have limitations that we address in this paper.
Heuristics-based approaches \cite{advisor,background_music} only consider the general mood of the music video and the user listening history. The mood categories are annotated independently for the two modalities, require manual annotations for every video and audio segment, and are restricted to a limited number of pre-defined discrete categories. %Our approach is not heuristics-based and it can learn the correspondence directly from data, and is not limited to mood correspondence.

Cross-modal ranking losses for music and video \cite{Li2019QueryBV,CBVMR,cca} and learned audio features \cite{pretet_cross-modal_2021} have been used to obtain state-of-the-art (SoA) results. We build on top of them with three key contributions that lead to a tenfold improvement in retrieval accuracy: we 1) propose a framework that models temporal context, 2) use a noise-contrastive loss \cite{oord2018representation}, which has been shown to be better suited to self-supervised settings, and 3) use SoA feature extraction models.

% The closest in methodology to our approach are  \cite{Li2019QueryBV,CBVMR,pretet_cross-modal_2021,cca}. 
% \cite{Li2019QueryBV} uses a cross-modal ranking loss, %To avoid losing modality-specific characteristics, they introduce a soft within-modality loss that leverages the relative distance relationship between intra-modal samples before embedding. 
% \cite{CBVMR} trains cross-modal embeddings with emotion tags as supervision, and
% \cite{pretet_cross-modal_2021} builds on \cite{CBVMR} and leverages learned audio features. 
% Finally, \cite{cca} aligns the music and video modalities by clustering their representations. There are three key differences between our approach and this prior work. First, our framework is capable of using temporal context, that allows us to model temporal evolution both in music and in video. Second, we use a noise-contrastive loss \cite{oord2018representation}, which has been shown to be better suited to self-supervised settings than other metric losses. And finally, we use state-of-the-art feature extraction models. Moreover, in \cite{cca} they also use supervised emotion information, which does not scale to large amounts of data. Overall, we improve previous state-of-the-art retrieval results by a large margin (see Section~\ref{sec:experiments}).

Music synthesis, a task that is hard on its own, can also be conditioned on a given input video. Approaches like generating MIDI files by looking at fingers playing \cite{su_audeo_2020,gan_foley_2020}, and directly generating sound (foley) using spectrograms \cite{ghose_autofoley_2020} have been proposed, but they can only exploit low-level signals that do not capture any artistic aspect of the video. 
% A different approach for music recommendation consists in generating the music, given a video. Generating MIDI files by looking at fingers playing \cite{su_audeo_2020,gan_foley_2020} can only exploit the low-level movement signal, and does not capture any artistic aspect of the video. Directly generating sound (foley) using spectrograms \cite{ghose_autofoley_2020} is very hard to generalize to music, and also limited to low-level and specific audio outputs, far from our artistic goal. Music synthesis is a hard task on its own, even without conditioning on video.
Using pre-trained music generation models, conditioning them on video \cite{dassani_automated_2019}, limits the audiovisual correspondence to a few pre-determined parameters (\emph{e.g.,} energy, direction, and slope), which cannot be learned in a self-supervised fashion.
% Other works use pre-trained music generation models, and use video to condition them \cite{dassani_automated_2019}. This approach is limited to the input parameters the music generation models admit (\emph{e.g.,} energy, direction, and slope). Also, the generation model is not retrained, which implies that the video parameters have to be either extracted heuristically, or learned in a supervised manner. 

Recent previous work \cite{pretet2021there} studied the relationship between some music traits (\eg, beats) and video editing operations (\eg, cuts) by interviewing professional editors and computing statistics on existing video data. They observe that some of the correspondences require contextual information; for example, some editors increase cutting video to the beat close to a climax moment in the video, or choose the video content to emphasize a musical climax. Such a finding {\em suggests} that there could be value in modeling temporal context for correspondence learning. Our paper {\em shows} that context is important quantitatively, while also showing how to best achieve this technically.

% \textbf{Long-term video understanding}
\textbf{Long-form video}
%Overall, a key aspect that all the existing literature on music recommendation from video has in common is that long-term video understanding is not considered. In our work, it has a pivotal importance. 
has been modeled in the literature by first computing representations at different temporal locations, and then combining them, either through averaging or by learning a more complex combination of temporal features \cite{NonLocal2018,wu_long-term_2019,TSN2016ECCV}. 
We propose to model the long-term temporal context in both the music and video modalities using Transformers \cite{Vaswani2017}, which use attention to model long sequences, and have become the SoA method for many NLP tasks in the past few years. %Their key property is that they can model the context in long sequences, using self-attention. 
Recently, they have been adapted to other domains such as images \cite{caron_emerging_2021,dosovitskiy_image_2020,fan_multiscale_2021,liu_swin_2021,chen_empirical_2021}, videos \cite{Bertasius2021,arnab_vivit_2021,akbari_vatt_2021,neimark_video_2021,bain_frozen_2021,wang_long-short_2021,wu2021towards}, audio \cite{verma_audio_2021,gong_ast_2021}, multi-modality inputs \cite{gabeur_multi-modal_2020}, and even modality-agnostic Transformers have been proposed \cite{jaegle_perceiver_2021,akbari_vatt_2021}, all with significant success.  
Video Transformers take as inputs either pixels directly, or features from pre-trained networks (\eg, \cite{neimark_video_2021}). We build on top of the latter approach, and experiment with both convolutional \cite{feichtenhofer_x3d_2020,radford_learning_2020,deepsim} and Transformer-based \cite{Bertasius2021} base features, both for the visual and music modalities.

\textbf{Audiovisual self-supervised learning} has been studied %music-text matching in Won \etal \cite{won2021emotion}, and for audio-visual 
in a number of papers \cite{alayrac_self-supervised_2020,Alwassel2019,yang_telling_2020,Arandjelovic2017a,Owens2015,Afouras2020,Suris2018Crossmodal} that deal with \textit{physical} events and sounds, such as the sound of dogs, cars, or musical instruments, or the location of the sound sources. However, these papers do not deal with the higher-level artistic \textit{music}-video correspondence. 
 \begin{figure}[t]
\centering
% \vspace{-0.2cm}
\includegraphics[width=\columnwidth]{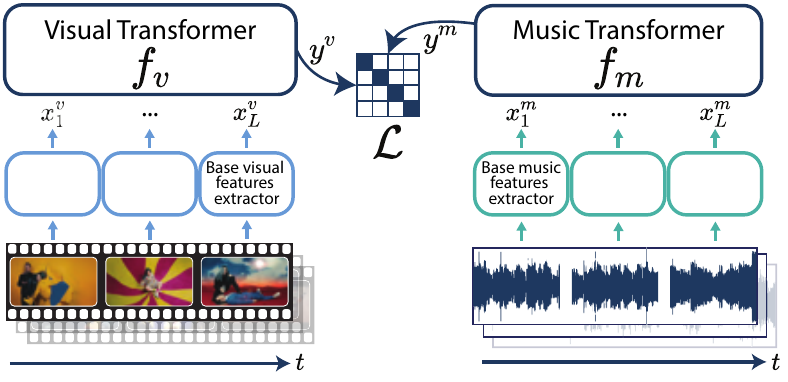}
\vspace{-0.5cm}
\caption{\textbf{Method}. We split music videos into visual and musical segments, pre-compute strong modality-specific base features, and process them separately using contextual Transformers. We self-supervise the model using an InfoNCE loss.}
\vspace{-0.5cm}
\label{fig:method}
\end{figure}
 
\textbf{Music-conditioned video editing} has mostly focused
%The flip side of recommending music for video is using music for video retrieval or generation. Our training approach is symmetrical for video and music, which makes it adaptable to this task, while keeping the artistic correspondence. All the work done in this domain focuses 
on synchronizing music with \emph{dancing} videos. Approaches range from video resampling to fit the music \cite{davis2018beat} to directly generating pixels of people dancing, conditioned on static images and a music track (\emph{e.g.} \cite{lee2019dancing2music,ren_mm_dance,ferreira2020cag}).
The focus on the dancing, while musically oriented, relies on low-level correspondence between music beats and human movement. This paper focuses on higher level correspondence, where emotions, story, and context are key factors, and are not considered in dancing videos.

% \DS{Not sure how to cite \cite{pretet2021there}}
% \JS{see my comment on slack :)}

% \bryan{May want to have a paragraph on fairness prior work.}\DS{I would leave that for the fairness section. Otherwise it would look like the paper is about fairness. Also, if we mention everything in one place it may be easier to have a compact explanation.}\JS{DS - sounds good to me.}

\section{Music Video Pretraining over time (MVPt)}
\label{sec:method}

In the following sections we describe our proposed method, Music Video Pretraining over \emph{time}, or MVPt.

% \subsection{Inputs and outputs}
% \label{sec:inputs}
\textbf{Inputs and outputs}.
During training, the inputs to the framework are a collection of video and music pairs $\videocorpus$ (\textit{music videos}), where the music and video have been artistically paired by human creators. Each raw music video is processed to obtain base representations for the {\em visual track} $\video$ and a {\em music track} $\music$. Additionally, each music video is divided into $\numsegments$ {\em segments}, of duration $t$. Correspondingly, these segments consist of a {\em visual segment} and a {\em music segment}. The division into segments allows us to 1) process the music video as a sequence, and thus exploit temporal context, and 2) make separate predictions for every segment, at a more fine-grained temporal resolution. %See Figure \ref{fig:inputs} for an illustration.

Our model takes $\video$ and $\music$ as inputs, and it outputs representations $y^v=f_v(x^v)$ and $y^m=f_m(x^m)$, respectively, where $f(.; \theta)$  represent the functions whose parameters $\theta$ are optimized. See Fig.~\ref{fig:method} for an overview of the framework.

\begin{table*}[t]
\small
\setlength\extrarowheight{-3pt}
\centering
\caption{\label{tab:ablations_clip}\textbf{Segment-level retrieval results for MusicVid-YT8M}. Each one of our contributions improves the accuracy of the model.}
\vspace{-0.2cm}
\begin{tabular}{@{}r@{\hspace{0.7\tabcolsep}}lrr rrr rrr r}
\toprule
{}&{}                                              & \multicolumn{2}{c}{Median Rank ${\downarrow}$} &   \multicolumn{7}{c}{Recall ${\uparrow}$} \\
\cmidrule(lr){3-4}
\cmidrule(lr){5-11}

{}&{}                                              & V$\rightarrow$M & M$\rightarrow$V &  \multicolumn{3}{c}{ V$\rightarrow$M} & \multicolumn{3}{c}{M$\rightarrow$V} & Average \\
\cmidrule(lr){5-7}
\cmidrule(lr){8-10}
{}&{}                                              &  &  &     R@1 &     R@5 &    R@10 &   R@1 &   R@5 &  R@10 & R@10 \\

\midrule
\scriptsize{1}&Baseline                                        & 349 & 277 & 0.55 & 2.16 & 3.83 & 0.74 & 2.79 & 4.83 & 4.33\tikzmark{a}  \\
\scriptsize{2}&\quad+ CLIP and DeepSim features                & 176 & 107 & 3.06 & 6.31 & 9.69 & 4.71 & 8.14 & 12.14 & 10.91  \\
\scriptsize{3}&\quad\quad+ Transformers (music time)      & 27 & 26 & 16.53 & 27.04 & 37.14 & 16.50 & 26.86 & 37.17 & 37.15  \\
\scriptsize{4}&\quad\quad+ Transformers (visual time)      & 24 & 24 & 17.23 & 27.54 & 38.64 & 17.07 & 26.85 & 38.43 & 38.54  \\
\scriptsize{5}&\quad\quad\quad+ InfoNCE (MVPt, ours)        & \textbf{19} & \textbf{12} & 17.33 & 29.12 & 39.33 & 19.98 & 34.81 & 45.41 & \textbf{42.37}\tikzmark{b}  \\
\midrule
\scriptsize{6}&MVPt + X3D features                     & 28 & 27 & 8.47 & 19.66 & 28.87 & 8.83 & 19.88 & 29.20 & 29.03  \\
\scriptsize{7}&MVPt + TimeSformer features             & 40 & 36 & 6.81 & 16.87 & 25.83 & 7.36 & 17.51 & 26.34 & 26.09  \\
\midrule
\scriptsize{8}&MVPt + $t=4.5$s              & 52 & 52 & 11.70 & 17.18 & 24.79 & 10.62 & 16.60 & 24.55 & 24.67 \\
\scriptsize{9}&MVPt + $t=11$s              & 7 & 6 & 28.97 & 47.71 & 65.82 & 29.37 & 47.52 & 65.32 & 65.57 \\
\midrule
&Chance                                          & 1000 & 1000 &	0.05 & 0.25 & 0.50 & 0.05 & 0.25 & 0.50 & 0.50 \\
\bottomrule
\end{tabular}
\begin{tikzpicture}[overlay, remember picture, shorten >=.5pt, shorten <=.5pt, transform canvas={yshift=.35\baselineskip, xshift=0.2\baselineskip}]
    \draw [->] ({pic cs:a}) [bend left] to ({pic cs:b});
\end{tikzpicture}
\end{table*}

\begin{table*}[t]
\small
\setlength\extrarowheight{-3pt}
\centering
\caption{\label{tab:ablations_video}\textbf{Track-level retrieval results for MusicVid-YT8M}. Each one of our contributions improves the accuracy of the model.}
\vspace{-0.2cm}
\begin{tabular}{@{}r@{\hspace{0.7\tabcolsep}}lrr rrr rrr r}
\toprule
&{}                                              & \multicolumn{2}{c}{Median Rank ${\downarrow}$} &   \multicolumn{7}{c}{Recall ${\uparrow}$} \\
\cmidrule(lr){3-4}
\cmidrule(lr){5-11}

&{}                                              & V$\rightarrow$M & M$\rightarrow$V &  \multicolumn{3}{c}{ V$\rightarrow$M} & \multicolumn{3}{c}{M$\rightarrow$V} & Average \\
\cmidrule(lr){5-7}
\cmidrule(lr){8-10}
&{}                                              &  &  &     R@1 &     R@5 &    R@10 &   R@1 &   R@5 &  R@10 & R@10 \\

\midrule
\scriptsize{1}&Baseline                                        & 234 & 98 & 0.76 & 3.42 & 5.90 & 2.57 & 8.61 & 13.81 & 9.86\tikzmark{c} \\
\scriptsize{2}&\quad+ DeepSim audio features                         & 142 & 94 & 1.41 & 5.23 & 9.01 & 2.29 & 8.55 & 13.80 & 11.41 \\
\scriptsize{3}&\quad\quad+ CLIP visual features                       & 64 & 45 & 3.00 & 11.13 & 18.56 & 5.03 & 15.70 & 24.09 & 21.33 \\
\scriptsize{4}&\quad\quad\quad+ Transformers w/o position      & 24 & 21 & 5.09 & 19.58 & 32.40 & 5.86 & 21.81 & 35.70 & 34.05  \\
\scriptsize{5}&\quad\quad\quad\quad+ Temporal embeddings       & 18 & 17 & 5.99 & 23.20 & 38.33 & 6.22 & 24.43 & 40.68 & 39.50 \\
\scriptsize{6}&\quad\quad\quad\quad\quad+ InfoNCE (MVPt, ours)   & \textbf{13} & \textbf{13} & 6.09 & 24.91 & 41.89 & 6.36 & 25.73 & 42.65 & \textbf{42.27}\tikzmark{d} \\
\midrule
{}&Chance                                          & 1000 & 1000 &	0.05 & 0.25 & 0.50 & 0.05 & 0.25 & 0.50 & 0.50 \\
\bottomrule
\end{tabular}
\begin{tikzpicture}[overlay, remember picture, shorten >=.5pt, shorten <=.5pt, transform canvas={yshift=.35\baselineskip, xshift=0.2\baselineskip}]
    \draw [->] ({pic cs:c}) [bend left] to ({pic cs:d});
\end{tikzpicture}
\end{table*}
% \vspace{-0.2cm}
\begin{table*}[t]
\small
\setlength\extrarowheight{-3pt}

\centering
\caption{\label{tab:results_movie}\textbf{Segment-level results for MovieClips}. Our contributions are also useful in movies, and are not specific to music video clips.}

\begin{tabular}{@{}r@{\hspace{0.7\tabcolsep}}lrr rrr rrr r}
\toprule
&{}                                              & \multicolumn{2}{c}{Median Rank ${\downarrow}$} &   \multicolumn{7}{c}{Recall ${\uparrow}$} \\
\cmidrule(lr){3-4}
\cmidrule(lr){5-11}

&{}                                              & V$\rightarrow$M & M$\rightarrow$V &  \multicolumn{3}{c}{ V$\rightarrow$M} & \multicolumn{3}{c}{M$\rightarrow$V} & Average \\
\cmidrule(lr){5-7}
\cmidrule(lr){8-10}
&{}                                  &  &  &     R@1 &     R@5 &    R@10 &   R@1 &   R@5 &  R@10 & R@10 \\

\midrule
\scriptsize{1}&Baseline + DeepSim + CLIP           & 189 & 128 & 2.1 & 5.8 & 9.36 & 2.94 & 8.48 & 13.34 & 11.35 \\
\scriptsize{2}&Baseline + DeepSim + CLIP + InfoNCE   & 74 & 58 & 2.53 & 7.95 & 14.99 & 4.05 & 12.93 & 23.85 & 19.42 \\
\scriptsize{3}&MVPt (ours)                      & \textbf{21} & \textbf{21} & 15.08 & 25.55 & 36.25 & 14.99 & 25.94 & 36.87 & \textbf{36.56} \\
\midrule
\scriptsize{4}&MVPt + X3D features         & 28 & 28 & 8.58 & 19.08 & 28.52 & 8.90 & 19.74 & 29.69 & 29.11 \\
\midrule
&Chance                              & 1000 & 1000 &	0.05 & 0.25 & 0.50 & 0.05 & 0.25 & 0.50 & 0.50 \\
\bottomrule
\end{tabular}
\end{table*}

% \label{sec:loss}
\textbf{Cross-modal self-supervision}.
The music and visual tracks in videos have a strong correspondence. The music that plays on top of the video is artistically related to the content of the video. We exploit this alignment as supervision: given a representation of a visual segment, our model is trained to predict the representation of the corresponding music segment, and vice versa.

The energy function we optimize computes a similarity between the representations of video and music segments, and encourages positive (corresponding) pairs to have a high similarity value, and negative (non-corresponding) pairs to have a low similarity value. In practice, this is implemented using the InfoNCE contrastive loss \cite{oord2018representation}:
{\small
\begin{equation}
\vspace{-0.1cm}
    \mathcal{L}_{v\rightarrow{}m} = - \sum_i^\videocorpus\sum_\indexsegment^\numsegments\left[\log\frac{\exp\left[{s(\outputvideo_{i,\indexsegment}},\outputmusic_{i,\indexsegment})/\tau\right]}{{\sum_{j}^\videocorpus\sum_\indexsegment^\numsegments\exp{\left[s(\outputvideo_{i,\indexsegment},\outputmusic_{j,\indexsegment})/\tau\right]}}}\right],
    \label{eq:contrastive}
\end{equation}}
% where $f_v(.)$ and $f_m(.)$ represent the visual and music networks respectively, and 
where $s(y^v, y^m)$ is the similarity function, which following common practice we implement as the cosine similarity $s(y^v, y^m) = \frac{{y^v}^Ty^m}{||y^v||\cdot ||y^m||}$. $\tau$ is a hyperparameter that we set to $\tau=0.3$, following \cite{chen2020simple}. 
$\mathcal{L}_{m\rightarrow{}v}$ is defined symmetrically, and the final loss is $\mathcal{L} = \mathcal{L}_{v\rightarrow{}m} + \mathcal{L}_{m\rightarrow{}v}$, which is used to train the model using stochastic gradient descent.

% Equation~\ref{eq:contrastive} shows a contrastive loss where the normalization is with respect to all the negative music segments, given a video segment. A symmetric loss where the normalization is with respect to all the negative video segment, given a music segment, is also used \DS{Should I explicitly show the other direction?}. The two losses are averaged, and the model is trained using stochastic gradient descent. \JS{you could update eq (1) to $\mathcal{L}_{v\rightarrow{}m} = ...$ and then add an eq 2: $\mathcal{L} = \mathcal{L}_{v\rightarrow{}m} + \mathcal{L}_{m\rightarrow{}v}$. It's kinda obvious but if we have space would be nice for completeness.}

% \subsection{Contextual model}
% \label{sec:model}
\textbf{Contextual models $f_v$ and $f_m$}.
Music and video are not only signals with a strong temporal component, they are also synchronized: changes in one modality are temporally aligned with changes in the other modality. Therefore, temporal context heavily impacts audiovisual correspondence, and needs to be modeled accordingly. To do so, we use a Transformer network \cite{Vaswani2017}, whose attention mechanism computes how much each element of the sequence has to attend to every other element in the sequence. We append a [CLS] token to the input, to represent the full video.

\textbf{Base features}.
In our experiments, we use deep pre-computed base features that are obtained from the visual and music raw signals. This allows us to 1) build upon state-of-the-art models and leverage large-scale pretraining, and 2) lift the representation demands from the Transformer networks, allowing them to focus their representation power on modeling the temporal context and the cross-modal alignment. Specifically, we use CLIP \cite{radford_learning_2020} for visual features and disentangled music tagging embeddings (DeepSim) \cite{deepsim} for music features. We temporally average the base features extracted for every segment of duration $t$.

% \subsection{Transfer Learning}
% \label{sec:transfer}
% \DS{Downplay importance. We don't want to run baselines with raw inputs}
% The signals $x^m$ and $x^v$ used in our framework represent the music and video signals, respectively. Our approach is general, and the signals $x$ can be represented using either raw signals (pixels in video and amplitude values in music), or features representing them. In our experiments, we use deep pre-computed embeddings that are obtained from these raw signals. This has two main advantages. First, it allows us to build upon state-of-the-art models, and leverage large-scale pretraining. And second, it lifts representation demands from the Transformer networks, allowing them to focus their representation power only on the temporal context and cross-modal alignment, which are the truly significant aspects of the framework. % This makes training easier and allows for faster iterations. 

% In our experiments we experiment with TimeSformer \cite{Bertasius2021}, X3D \cite{feichtenhofer2020x3d} and CLIP \cite{radford_learning_2020} features for video, and DeepSim \cite{deepsim} and OpenL3 \cite{cramer2019look} for music. Our framework is not restricted to these choices and can work on top of any pre-computed features, as well as raw inputs.

% \subsection{Inference}

\textbf{Inference}.
At inference time, the model takes a video as input, splits it into segments, and computes contextualized features for all segments. For each visual segment, it recommends a music segment that matches both the content of the visual segment, as well as the contextual information around it. The music segments are selected from a pool containing all the available music segments in the (test) dataset, according to the similarity metric used during training. The music to video retrieval is done equivalently.
See Appx.~\ref{sec:app_implementation} for implementation details.

\section{Retrieval Experiments}
\label{sec:experiments}

% \subsection{Datasets}
% \label{sec:datasets}
We show retrieval experiments in two different settings. In the first setting (``track level''), we retrieve an entire full-length music audio track given a full-length query video (and vice versa). This setting allows evaluating the quality of the retrievals at the level of an entire (untrimmed) video. In the second setting (``segment level''), we aim to evaluate a finer-grained alignment between the two modalities where we retrieve a short music segment given a short video segment. In both the segment- and track-level settings, the inputs to our model are the $L=30$ segments comprising the complete music video. %For the track-level setting, we take the output of the [CLS] as the representation of the full video.
% (You need to say here how you incorporate the full video/audio track as context for your method and the baselines)

Given a query visual track (or music track), we compute the feature distance to each music track (or video track) in a pool of $\targetsize$ target candidates not seen during model training, where we set $\targetsize=2000$ in all of the experiments following the setup in Pr\'etet \etal~\cite{pretet_cross-modal_2021}. Only one of the candidates is the correct pair (ground truth). We then sort the candidates according to this distance value and use two different criteria %metrics
to evaluate the success of the retrieval. \textbf{Recall@$K$ {\normalfont(the higher the better)}}: 
we look at the $K$ closest candidates and consider the retrieval successful if the ground truth pair is among those, and we report the percentage of successful retrievals in the test set. \textbf{Median Rank {\normalfont(the lower the better)}}: we return the position of the ground truth pair in the sorted list of candidates; we then report the median of the position values across the test set.
% \end{itemize}

Our approach is general and adaptable to any video data that contains music. We evaluate our method on two different datasets. \textbf{YT8M-MusicVideo}: we leverage a set of 100k videos from the YouTube8M dataset \cite{abu2016youtube} that are tagged as ``music video'', with an average track duration of 4 minutes. We use segments of duration $t=6.7s$ for this dataset. %Music videos contain a strong signal relating music to its visual content, which makes them a natural fit for our framework.
% The average duration of the music videos in the dataset is 4 minutes. 
\textbf{MovieClips}:
we collect all videos from the MovieClips YouTube channel \cite{movieclips}. %\bryan{Add details here on how you choose which videos to collect.} 
From these videos, we select the parts that contain music consecutively for at least 20s. We did so by training a PANN model \cite{9229505} on AudioSet \cite{gemmeke2017audio} and used it to detect regions with music in the data. The final number of selected video tracks in the dataset is 20k and their average duration is 42 seconds, and we use $t=3.3s$. %The alignment between movies and music requires a stronger representation of emotions compared to music video clips, as it is the main signal that is available for correspondence. \bryan{$\leftarrow$ Do we need this last sentence? We mention it a lot already in the intro.}
See Appx.~\ref{sec:app_datasets} for more information about the datasets' statistics and creation process.
\definecolor{mycolor}{HTML}{143a61}
\definecolor{mycolor2}{HTML}{FF5733}
\begin{figure}
\centering
\vspace{-0.1cm}
\begin{tikzpicture}
    \pgfplotsset{%
        width=1.\columnwidth,
        height=0.5\columnwidth
    }
    \begin{axis}[
        legend style={at={(0.02,0.78)},anchor=west},
        xlabel=Sequence length (seconds),
        ylabel=Recall@10,
        xmin=6.66, xmax=200,
        ymin=0, ymax=50,
        % xtick={10,20,30},
        % xticklabels={A,B,X},   % <---
        ytick={0,10,...,100},
        every axis plot/.append style={ultra thick},
        grid = both,
                ]

    \addplot[smooth,mycolor] plot coordinates {
        (1*6.66, 7.51)
        (3*6.66, 11.62)
        (5*6.66, 14.37)
        (7*6.66, 18.44)
        (9*6.66, 20.55)
        (10*6.66, 21.30)
        (11*6.66, 21.97)
        (13*6.66, 22.70)
        (15*6.66, 23.69)
        (17*6.66, 25.53)
        (19*6.66, 27.70)
        (21*6.66, 29.20)
        (22*6.66, 31.67)
        (23*6.66, 33.72)
        (25*6.66, 37.23)
        (27*6.66, 39.81)
        (28*6.66, 40.32)
        (29*6.66, 41.88)
        (30*6.66, 42.27)
    };
    \addlegendentry{Transformer}
    \addplot[smooth,mycolor2] plot coordinates {
        (1*6.66, 10.77)
        (2*6.66, 12.67)
        (4*6.66, 15.81)
        (7*6.66,17.39)
        (9*6.66,18.23)
        (12*6.66, 19.03)
        (14*6.66, 19.49)
        (19*6.66, 20.13)
        (24*6.66, 21.07)
        (30*6.66, 21.33)
    };
    \addlegendentry{MLP}

    \end{axis}
\end{tikzpicture}
\vspace{-0.4cm}
\caption{\textbf{Temporal context}. The accuracy of the transformer decreases as we remove temporal context, indicating its importance.} \label{fig:temporal_context}
\end{figure}
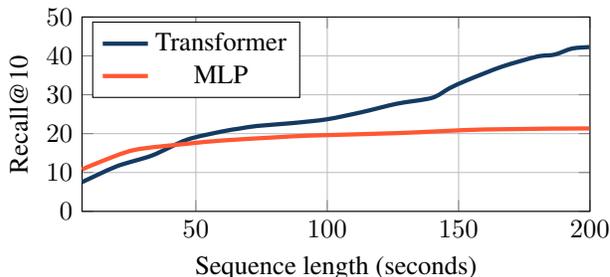

\textbf{Baseline}.
% \bryan{You may want to have this paragraph come after the evaluation criteria and task definition, where the reader will have full context for the baseline (see my comment below on the Pretet baseline).}
We build our contributions on top of the prior SoA method of Pr\'etet \etal~\cite{pretet_cross-modal_2021}. They propose a similar framework, but train with a triplet loss instead of an InfoNCE loss, use an MLP model instead of a Transformer, and use ImageNet base features for video and OpenL3 \cite{openl3} base features for music. We refer to this model as ``Baseline''. The input to the baseline model is the average across time of all the base features for the track level setting, and the average of the base features over a single segment for the segment-level setting. Our re-implementation of the baseline yields a retrieval accuracy (9.86\% track level Recall@10) on our test set, that is close to the results reported by Pr\'etet \etal (12.10\%) for the MVD dataset \cite{mvd}, which we did not have access to. MVD is a manually curated subsample of the YT8M-MusicVideo dataset, so similar (same parent dataset) but slightly better (curated for clean audiovisual correspondences) results are to be expected. 
% \bryan{You may want to note that your re-implementation of the baseline yields a retrieval accuracy that is close to the reported results of Pretet \etal. (quote their number, say that your results are in line with theirs, and perhaps explain what's different in your setup from theirs).} 

\textbf{Ablations}.
In our results, we show how modifying each of the model components %one of the previous options 
contributes to an increase in retrieval accuracy.
Note that 1) the MLP baseline has access to the same set of base features for each modality, but the features are aggregated via an average-pool operation before being passed as input to the MLP, and 2) in our Transformer model we match the number of model parameters (5.5M) to the baseline MLP, so the model capacity is not an advantage of our method. 
%Note that 1) the MLP baseline has access to the same temporal context information, but \br{the features are} aggregated in a different way, and 2) in our Transformer model we match the number of parameters of the baseline MLP, so model capacity is not an advantage of our method. 
% \bryan{Say what are the MLP/Transformer model capacities.}
When using Transformers, a temporal encoding is added to each segment input. This setup allows the Transformer to exploit temporal information on top of the contextual one. In the segment-level setting, using temporal encodings for {\em both} modalities can result in a learning shortcut, where the model learns to associate visual segments to music segments based on their position in the sequence. Therefore, in the segment-level experiments, we disable the temporal embeddings for one of either the visual or the music modality. We report results for both options, indicating which modality keeps the temporal encoding as ``music time'' or ``visual time''. 
%\bryan{Which one?} 
Note that the Transformer is still capable of using contextual information. 

\textbf{Results}. We show segment-level results for YT8M-MusicVideo and MovieClips in Tab.~\ref{tab:ablations_clip} and \ref{tab:results_movie}, respectively, and track-level results for the YT8m-MusicVideo dataset in Tab.~\ref{tab:ablations_video}.
The results show how each one of our contributions improves the performance of the model. Specifically, the modeling of temporal context via a Transformer proves critical (rows 2 and 3 in Tab.~\ref{tab:ablations_clip} and row 4 in Tab.~\ref{tab:ablations_video}). Additionally, the results show that representing temporal information (on top of context) in the visual track (``visual time'' in Tab.~\ref{tab:ablations_clip}, row 4) is just slightly more beneficial than modeling time in the music track (``music time'', row 3).

Also, as shown in both Tab.~\ref{tab:ablations_clip} (rows 6 and 7) and \ref{tab:results_movie} (row 4), using video-trained base features like X3D \cite{feichtenhofer2020x3d} or TimeSformer \cite{gberta_2021_ICML} base features, while being better than the baseline ones, result in worse performance than using CLIP base features %ones 
(which are trained on images). We argue that the reason for this result is that CLIP has been trained to align images with natural language sentences with a large vocabulary on a {\em larger} and more {\em generic} corpus of data than the ImageNet, X3D and TimeSformer models, that have been trained on fixed-vocabulary classification tasks. 
Finally, using longer segments improves the segment-level retrieval, at the expense of having a less fine-grained temporal representation (row 9 in Tab.~\ref{tab:ablations_clip}).
% \JS{How many frames does our approach use? It's not mentioned anywhere so the reader has no context as to what 40 or 100 frames mean (is that more than ours? less?).}
% \bryan{Point out where we should look to see this result. Maybe number the rows of the tables so you can directly refer to the results.}

To study the importance of temporal context, we evaluate our model when the input sequences are shorter, given the same segment length $t$. As seen in Figure~\ref{fig:temporal_context}, the model's accuracy decreases as the sequence length decreases, which demonstrates the importance of temporal context. In comparison, the MLP version is not even capable of exploiting long temporal contexts in the first place.
% \bryan{If there is time, it may be good to show the same plot for the MLP so you can directly compare the two models.} \JS{It can be another curve (MLP) in the same plot}

Finally, to show that our approach captures correspondence from an \emph{artistic} level, we perform human experiments: given a query video (or music) segment, we ask humans to choose between a music (or video)
segment retrieved by our model, and another one retrieved by the baseline. 71.4\% of responses prefer our method over the baseline, validating our claim ($p$-value $<0.01$) that our model is preferable from an artistic viewpoint. More details are provided in Appx.~\ref{appx:human}.

% \subsubsection{Clip-level retrieval}
% \subsubsection{Importance of temporal context}
% \subsection{Controlling for Length Bias}
\begin{figure}[t]
\centering
\includegraphics[width=1\columnwidth]{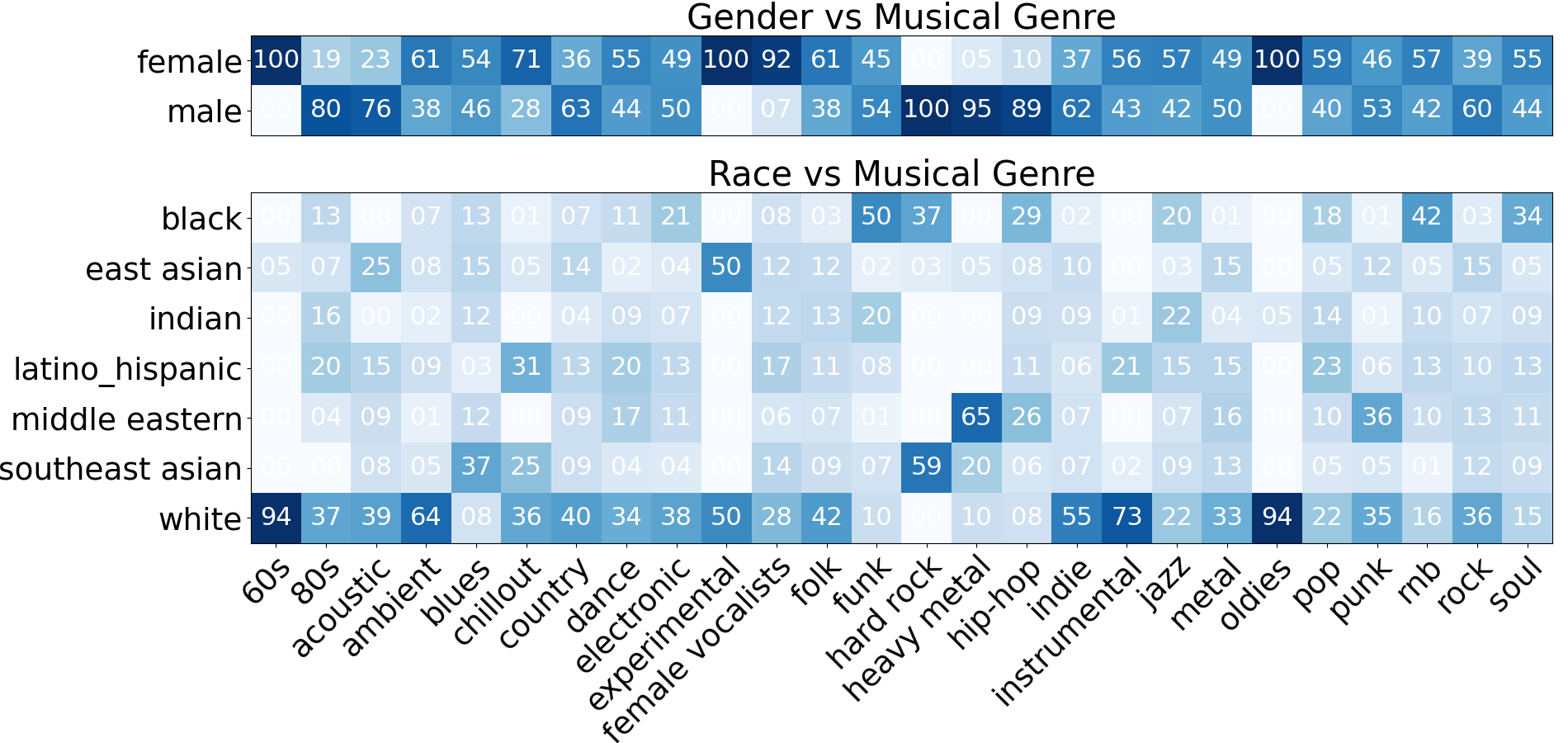}
\vspace{-0.5cm}
\caption{\textbf{Gender and race vs music genre}. For a given music segment with a genre annotation (not used during training), we retrieve the closest face image from the FairFace \cite{karkkainenfairface} dataset. We plot the gender and race of the retrieved image, normalized for every genre (each column adds up to 100). See Section~\ref{sec:fairness} for discussion about biases.}
\vspace{-0.5cm}
\label{fig:gender_race_genre}
\end{figure}

\section{Analysis and Applications}
\vspace{-0.2cm}
\label{sec:analysis_applications}

In this section, we probe what our music-video model has learned, showing that it learns to use a wide range of signals, from relatively low-level ones (like music tempo) to high-level ones (like music genre). Additionally, we qualitatively evaluate the retrieval soundness of our model and show that both retrieval directions return samples that match the query at a remarkable level, to the point that, ignoring lip-syncing, usually look and sound correct to the human eye (and ear). Finally, we show how we can condition the retrieval results to return samples that contain specific attributes, and visualize the attention in our model.

\definecolor{mycolor2}{HTML}{FF5733}
\begin{figure}[t]
\includegraphics[width=\columnwidth]{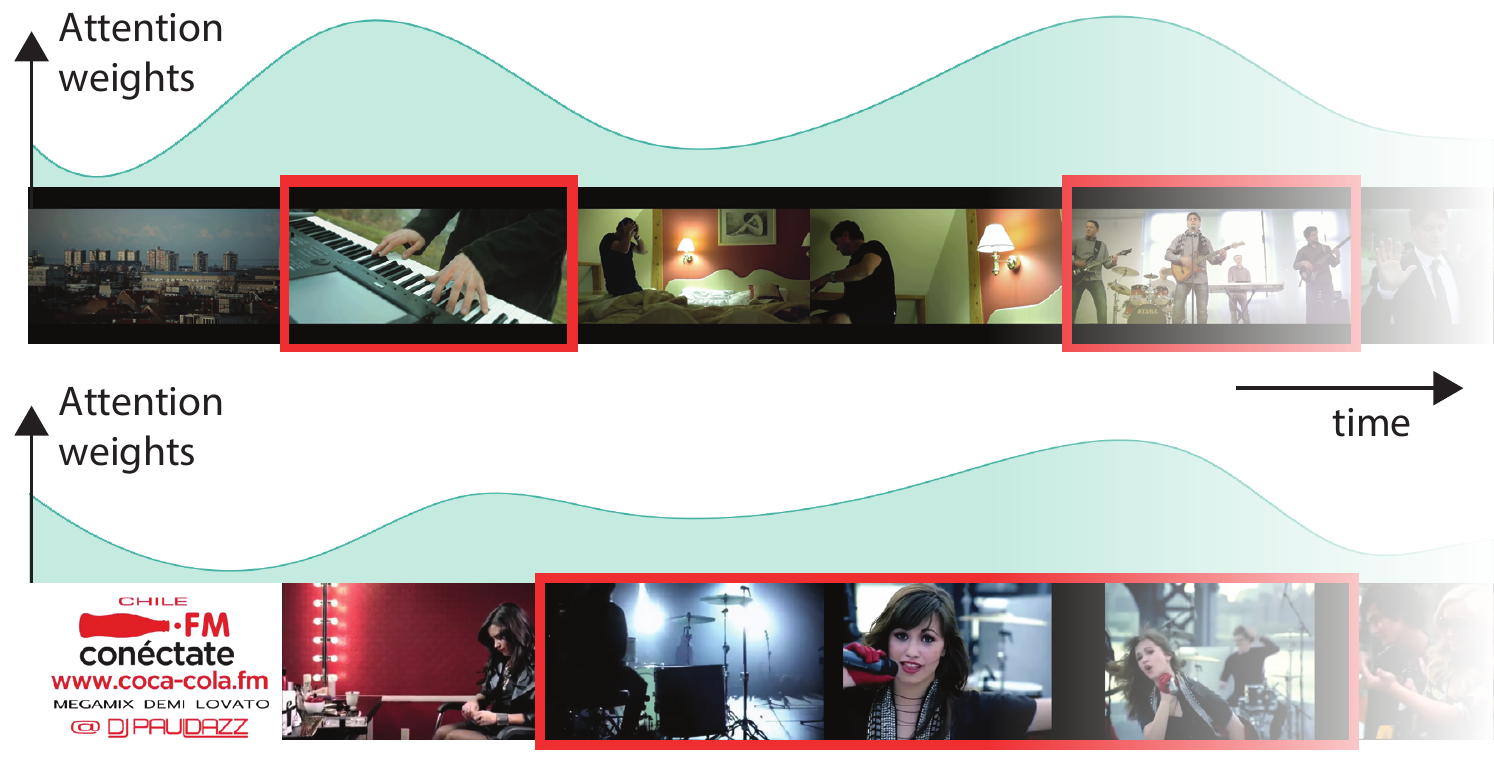}
\centering
\caption{\textbf{Visualization of attention}. We show the aggregation of the attention weights for every input segment, in two different examples (that we show partially). In every case, we \textcolor{mycolor2}{highlight} in red the segments that contain people singing or playing instruments. We notice that these correspond to the segments with high attention values, which implies the model prefers to use this information over less music-related moments.}
\vspace{-0.4cm}
\label{fig:attention}
\end{figure}

\subsection{Quantitative Analysis}
\label{sec:analysis}
We consider eight audio, visual, or audiovisual attributes: color brightness and hue, tempo, background scene, musical instruments, age, race, and gender. We include the latter two, in particular, to allow us to study questions related to bias. We adopt the definitions for race and gender from work targeting fairness in machine learning \cite{karkkainenfairface}. We study how each one of them influences our model's predictions. We perform all the analyses on the YT8M-MusicVideo dataset. %For a more detailed setup of our experiments, we refer the reader to Appx.~\ref{sec:app_experiments}.
See Appx.~\ref{sec:app_experiments} for more details.

\textbf{Color brightness and hue}.
For every frame in the test set, we modify its brightness by a factor of $r$. Then, we perform retrieval as explained in Section~\ref{sec:experiments} and compute the average Recall@10 % in \Figure{tempo_brightness} 
for different values of $r$. Surprisingly, the Recall@10 accuracy only decreases 1 point (42.37\%$\rightarrow$41.24\%) for a brightness variation of up to 30\% ($r\in\{0.7, 1.3\}$), so brightness does not play an important role in our model's accuracy, suggesting it may be using higher-level visual clues, which we analyze next. Likewise, we analyze the importance of hue: while more significant than brightness, hue is not crucial to model performance.

\textbf{Tempo}.
We time-stretch the query music signal by a factor of $r$ to modify its tempo (\ie, make it slower or faster). We then evaluate retrieval %as explained in Section~\ref{sec:experiments} 
and compute the average Recall@10 %in \Figure{tempo_brightness} 
for different values of $r$. When modifying the tempo at a rate of 30\% ($r\in\{0.7, 1.3\}$), Recall@10 accuracy drops more than 5 points (42.37\%$\rightarrow$36.96\%). 
We show the curves of Recall@10 as a function of $r$ for both brightness, hue, and tempo in Appx.~\ref{app:quantitative}.

The subsequent attributes are all evaluated in the same way. We select an image dataset that contains annotations about that attribute and is balanced across the attribute classes. %that it is balanced across the values that attribute can take. 
We compute representations for all the images in the dataset using the visual branch of our model and use them as target candidates. Note that our base CLIP features operate at the image level; we pass in a single base CLIP feature for the image to our Transformer model.
Then, we use the music segments in our test set as queries and return the top-1 retrieved image from the balanced dataset. Finally, we plot a matrix of (music genre)-(visual attribute), normalized for every music genre. The music genre annotations, collected using musicnn \cite{pons2019musicnn}, are {\em only} used for analysis, not during training. Note that because the target retrieval (image) dataset is balanced, the preference for each one of the attribute values is fully determined by the model.

\textbf{Gender}.
We use the FairFace dataset \cite{karkkainenfairface}, which contains images of faces, categorized by genre, race, and age. We show results in Figure~\ref{fig:gender_race_genre} and discuss
potential bias in Section~\ref{sec:fairness}. It is worth noting that a bias exists, and it is coherent with what we would expect in the real world---like female images being associated with the genre ``female vocalists''---and with the bias in the training data---like male images being associated with ``hip-hop'', ``hard rock'', or ``heavy metal''. Less than half of the genres show a strong preference for one of the genders, meaning that the model often does not rely on this attribute to reason about genre.

\textbf{Race}.
Using the FairFace dataset, we repeat the previous analysis and show results in Figure~\ref{fig:gender_race_genre}. As expected, 
given the observed bias in our training data (Appx.~\ref{sec:app_musicvideos}), ``hip-hop'' is mostly associated with black people, while ``country'' is mostly associated with white people. We attribute some unexpected associations to the lack of representation of certain races in our training dataset. 

\textbf{Age}. We found age is not as important for the model as other attributes, so we moved the analysis to Appx.~\ref{app:quantitative}. 

\begin{figure}[t]
\includegraphics[width=\columnwidth]{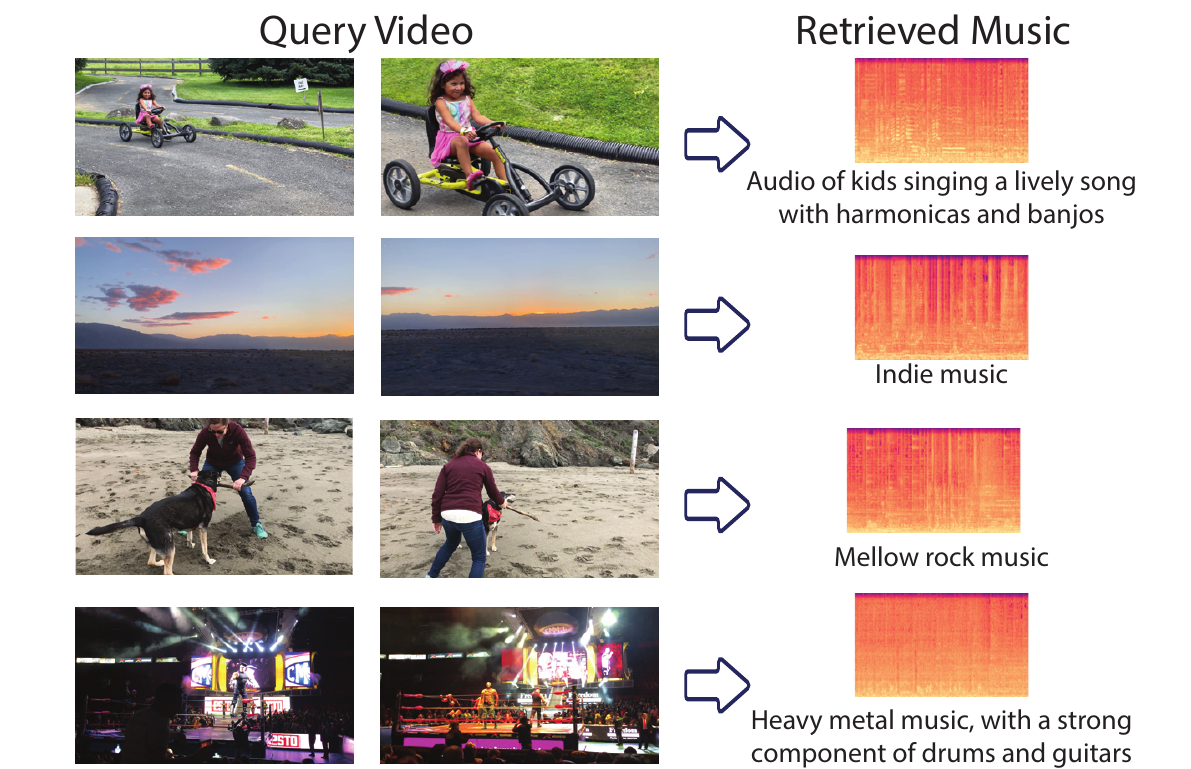}
\centering
\vspace{-0.4cm}
\caption{We test the YT8M-MusicVideo model on a set of casually captured videos outside of the dataset, and show how our model generalizes to scenes that do not naturally contain music.}
\vspace{-0.4cm}
\label{fig:homemade}
\end{figure}
\begin{figure}[t]
\includegraphics[width=1.0\columnwidth]{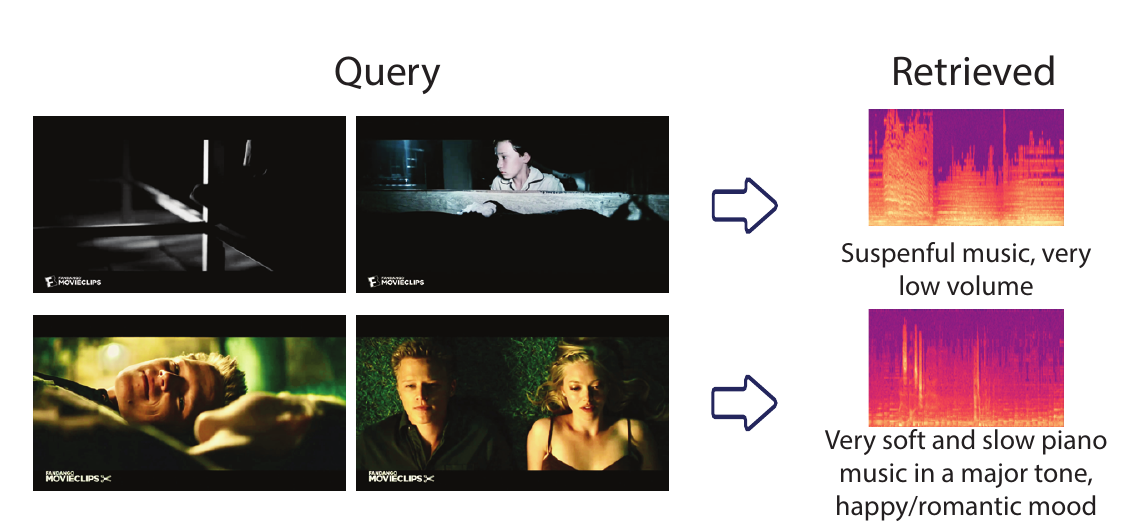}
\centering
\vspace{-0.6cm}
\caption{\textbf{Retrieval in the MovieClips dataset}. We show video-to-music retrieval examples, and show how our model exploits emotion to make the correspondence.}% See the Supp. Material for playable visualizations and more examples in both directions.}
\vspace{-0.5cm}
\label{fig:movies}
\end{figure}

\textbf{Visual Scene}.
We use the Places dataset \cite{zhou2017places}, which contains images of 205 scene categories. We observe that the scene attribute is also correlated with music genre, albeit less than the previous attributes. 
For reference, we list the most and least commonly retrieved scenes\footnote{Most commonly retrieved: \textit{track outdoor, runway, stage indoor, music-studio, boxing ring, baseball field, stadium baseball, martial arts gym, shoe shop, ballroom}; Least commonly retrieved: \textit{canyon, residential neighborhood, snowfield, arch, attic, desert vegetation, crevasse, fire escape, mausoleum, water tower}} and show the genre$\rightarrow$scene table in Appx.~\ref{app:quantitative}.

\begin{figure*}[h]
\includegraphics[width=0.95\textwidth]{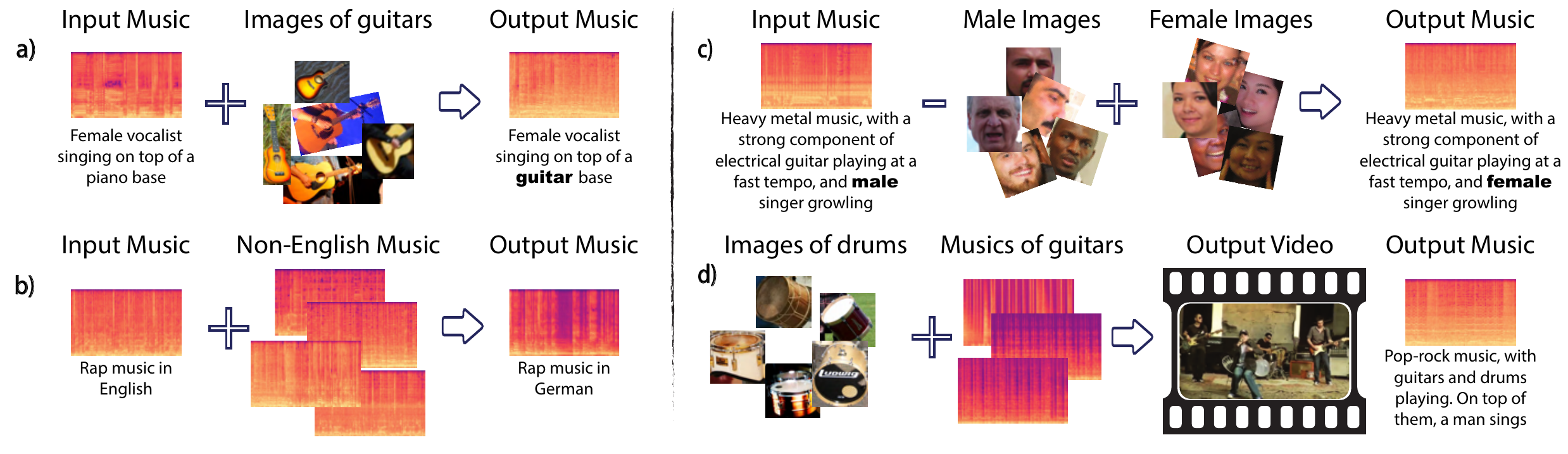}
\centering
\vspace{-0.1cm}
\caption{\textbf{Attribute conditioning}. Given data representing an attribute, like a set of guitar images to represent the ``instrument'' attribute, we can condition the retrieval of our model. Our model has {\em not} been trained with any attribute annotation. This conditioning is cross-modal, meaning that visual attributes can condition music outputs, and vice versa. Note that this figure is not a diagram of the procedure; we are showing actual examples. These operations can be consistently replicated in other examples, across musical genres. The inputs, output, and any of the conditioning attributes can be defined in any of the modalities.}
\vspace{-0.2cm}
\label{fig:neural_algebra}
\end{figure*}

\textbf{Visual Objects (Musical Instruments)}.
We use instrument images from the Open Images Dataset \cite{kuznetsova2018open}, and proceed as in the previous attribute studies. As exemplified by the attribute conditioning (description below) in Figure~\ref{fig:neural_algebra}a) and \ref{fig:neural_algebra}d), the model learns a strong and useful representation for some instruments (\eg, guitar, drums) in both the visual and musical modalities. The genre$\rightarrow$instrument table, shown in Appx.~\ref{app:quantitative}, shows a clear preference from most music genres to retrieve guitar images, especially for hard rock, heavy metal, metal, and punk genres. This is to be expected because 1) guitars are the most common instrument in these genres and 2) the video clips associated with these genres consist mainly of people playing the song, as opposed to other genres where the content is more cinematic.%The analysis reveals other interesting correlations, like ``funk'' retrieving with very high probability images of trombones, trumpets, and saxophones. 

\subsection{Qualitative Analysis and Applications}
\label{sec:qualitative}

\textbf{Retrieval examples}.
The best way to show the quality of our model is to put it to the test on real examples. We evaluate the model on visual and music segments obtained from YT8M-MusicVideo, shown in Figure~\ref{fig:musicvideo_retrieval}, and MovieClips, in Figure~\ref{fig:movies}. In Figure~\ref{fig:homemade}, we test the YT8M-MusicVideo-trained model on a set of casually captured videos outside of the YT8M-MusicVideo dataset and show how our model can generalize to scenes that do not naturally contain music.

\textbf{Attribute conditioning}. 
Knowing that our model captures a range of audiovisual attributes, we propose using their representations to condition the retrieval process. In order to find a representation $y_a$ of a specific attribute (\eg, guitar), we use an auxiliary dataset with labeled images and/or audios representing that attribute (\eg, images of guitars), compute their representations, and average them to obtain the representation of the attribute. We implement the conditioning by adding the representation $y$ of the query segment to the attribute one: $y_\text{conditioned} = y+y_a$. %There is no need for normalization as the cosine similarity is invariant to scale. 
If we instead want to \textit{remove} the attribute, we use a subtraction: $y_\text{conditioned}=y-y_a$. We can apply these operations 
multiple times, for attributes defined using either of the modalities. To deal with the potential out-of-distribution problem when conditioning on data from a different domain, we found that better results are obtained when instead of $y_a$ we use $y_a'=y_a-\sum_{b\in\mathcal{D}}y_b$, where $\mathcal{D}$ is the dataset the conditioning images or music tracks were obtained from (\eg, a dataset with images of instruments). 

Overall, this procedure gives rise to a variety of applications, ranging from video editing---where we want specific attributes to be present---to music or video search. We showcase this variety of approaches with some examples in Figure~\ref{fig:neural_algebra}. For instance, given an input music track not containing an instrument, we can retrieve a similar music track that contains that instrument. This instrument can be defined through data visually, as in example a), or via music, as in d). We can also condition on language. Specifically, in example b) we list a set of music tracks with non-English vocals, and we use them to retrieve music that is similar in style to a query input, but in a non-English language. Interestingly, the model creates a good representation of English (and as a consequence, non-English), but it is less consistent when representing other languages, probably due to the high proportion of English music tracks in the dataset.

\textbf{Attention}.
We visualize attention results from the visual Transformer $f_v$ in Figure~\ref{fig:attention}. For every example, we plot the attention weights at every video segment (represented by an image frame), computed using attention rollout \cite{abnar-zuidema-2020-quantifying}. 
These visualizations show that the model pays more attention to visual segments that contain people explicitly playing instruments or singing, over more cinematic content.

\section{Discussion and Limitations}
\label{sec:fairness}

% \DS{Any word more general than ``fairness''? This is not just about fairness I think.}
% \bryan{How about titling this section as ``Discussion and Limitations''?}

The correspondence between music and video is an artistic one. Art, and as an extension culture, is intrinsically tied to concepts such as language, nationality, gender, and race. Computer vision unfortunately still does not have the tools to deal with them in a satisfactory way. The result is a framework that resorts to bias and all the known negative effects bias can have in real-world applications \cite{howard2018ugly,suresh2019framework,mehrabi2021survey}. 

Unlike other recognition applications \cite{d2017conscientious}, however, in the context of artistic correspondence a framework that is invariant to these factors could lead to the erasure of cultural traits or to cultural appropriation. On the other hand, explicitly magnifying the ties between music and culture---as often done by the music industry \cite{ROY2004265}---can exacerbate certain biases or associations.

% These ties between music and culture have often been explicitly magnified by the music industry. As exposed by Roy \cite{ROY2004265}, the business logic in the 1930s in the U.S. resulted in an explicit racial coding of the music industry, whereby advertisements were targeted to different racial groups based on music genre. While this segregation eventually declined, the author notes that it did not disappear.

In this paper, we adopt a descriptive approach and present the correspondences the model is learning. We consider this paper an invitation for further study and discussion of the interplay between culture and bias in the context of artistic correspondence learning and the challenges it presents. These advances will require collaboration between computer science and sociology. The complex question of how to appropriately design such a system for real-world applications remains an open question. 

Finally, %while our human experiments suggest high-level concepts (such as aesthetics or overall story) may be captured, 
an explicit definition and precise evaluation of such concepts is lacking in our field and in this paper, and is an interesting avenue for future work.

%%%%%%%%% REFERENCES
{
    \clearpage
    \small
    \bibliographystyle{ieee_fullname}
    \bibliography{macros,main}
}

\clearpage
% --- supplementary material
\appendix

% --- PDF will be split by an editor (e.g. macOS preview), so need to restart from page 1
\setcounter{page}{1}

% --- repeat the title (AT: haven't found a more elegant way to do this...)
\twocolumn[
\centering
\Large
\textbf{Appendix} \\
% \vspace{0.5em}Supplementary Material - Appendix\\
\vspace{1.0em}
] %< twocolumn
\appendix

We divide the Appendix in three sections. First, in Appx.~\ref{sec:app_datasets} we describe the datasets in more detail. Second, in Appx.~\ref{sec:app_implementation} we provide more implementation details about the model. And finally, in Appx.~\ref{sec:app_experiments} we describe the experiments setup in more detail, and provide all the results that were not provided in the main paper. 

% \section{Qualitative Results}
% \label{sec:app_qualitative}

% See \href{http://musicvideo.cs.columbia.edu}{our website} for a video that show the results in Figures 1, 6, 7 and 8 of the main paper.
% The results show a query video or music track, and the goal is to retrieve a music track (or video) that pairs with it. Note: we exclude the audio track that originally paired with the query video (or vice versa) in all retrieval results to illustrate the learned artistic correspondences.

\section{Datasets Details}
\label{sec:app_datasets}

\subsection{YT8M-MusicVideos}
\label{sec:app_musicvideos}
In this section we give details about the order to study the distribution of music genres in our YT8M-MusicVideos dataset (in the music part), and of gender, race, and age (in its visual part). 

\begin{figure}[t]
\includegraphics[width=\columnwidth]{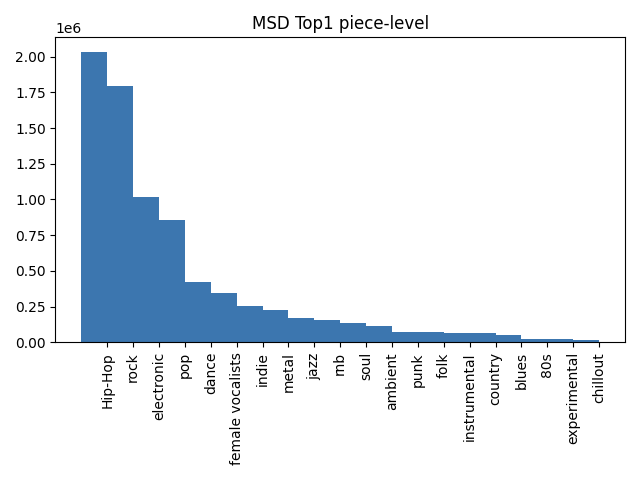}
\centering
\caption{Distribution of the top-20 genres in the YT8M-MusicVideos dataset.}
\label{fig:genres_dataset}
\end{figure}

We compute genre information using musicnn \cite{pons2019musicnn}. We predict a single category from the MSD dataset \cite{Bertin-Mahieux2011} for every music track, and show its distribution in Figure~\ref{fig:genres_dataset}.

\begin{figure}[t]
\includegraphics[width=\columnwidth]{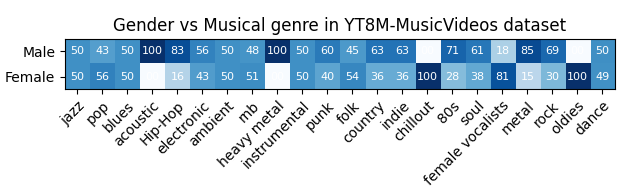}
\centering
\caption{Distribution of genders per music genre, normalized for every genre, in the YT8M-MusicVideos dataset. Note that this is {\em not} the same as Figure 4 in the main paper. Here we show dataset statistics, and in the main paper we show our model's retrieval results. Most of the very skewed distributions are skewed because those categories do not contain a lot of examples in the dataset (see Figure~\ref{fig:genres_dataset}). However, we note the very skewed (while representative) distribution of the hip-hop, female vocalists, and metal genres.}
\label{fig:gender_dataset}
\end{figure}

\begin{figure}[t]
\includegraphics[width=\columnwidth]{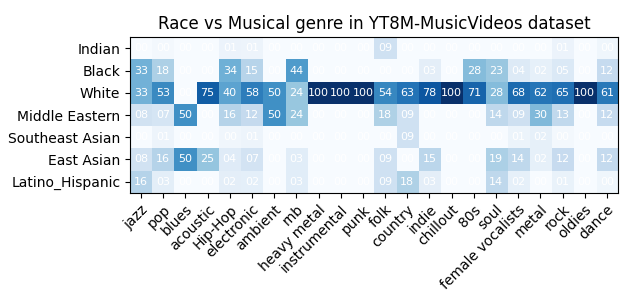}
\centering
\caption{Distribution of races per music genre, normalized for every genre, in the YT8M-MusicVideos dataset. Note that this is {\em not} the same as Figure 4 in the main paper. Here we show dataset statistics, and in the main paper we show our model's retrieval results.}
\label{fig:race_dataset}
\end{figure}

We study the gender, race, and age information using the FairFace dataset \cite{karkkainenfairface}. First, we collect a random subset of 2079 samples in our dataset, and we extract faces from the raw frames (we sample one frame per second in the selected videos) using a pre-trained face detection model. Second, we train a classification model on the FairFace dataset. Note that the FairFace dataset is uniform across races and genres, which means the trained model is not biased. However, we note that the test-time accuracy of our model is not perfect (72\% accuracy for race, 93\% for gender), which makes this analysis just orientative. Finally, we classify the previously extracted faces using this model, and represent each video with the most common category across all the faces extracted from it. If a video does not contain faces, we do not use it for the analysis. We show results in ~\Cref{fig:gender_dataset,fig:race_dataset}. In those figures, the frequencies are normalized for every gender. 

The distribution of races, genders and ages in the random subset of 2079 videos of our dataset is the following:
\begin{itemize}
    \item Gender: \{Male: 1368, Female: 711\}
    \item Race: \{Indian: 21, White: 1098, Middle Eastern: 283, Black: 414, East Asian: 192, Latino\_Hispanic: 57, Southeast Asian: 14\}
    \item Age:  \{ 0-2: 2,  3-9: 48, 10-19: 67, 20-29: 1644, 30-39: 287,  40-49: 17, 50-59: 10, 60-69: 4\}
\end{itemize}

\subsection{MovieClips}

In this section, we describe the procedure we used to create the MovieClips dataset. First, we downloaded all the videos from the YouTube channel MovieClips \cite{movieclips}, which contains 37k videos. From these videos, we selected the parts that contain music consecutively for at least 20s. We did so by training a PANN model \cite{9229505} on AudioSet \cite{gemmeke2017audio} and used it to detect regions with music in the data. Specifically, we consider a segment of audio a ``music'' one if any of the music-related classes in AudioSet is predicted with at least a probability of 0.3 (the probabilities for different classes are predicted independently). We consider music-related classes all the ones under the ``Music'' category in the AudioSet ontology. In order to filter out non-music audio, we additionally ignore the segments that (even though they may contain music) contain any other audio category with a probability of 0.2 or more. These predictions are done at a very small temporal resolution, resulting in small gaps of silence between music segments. In order to not treat these segments as separate, but as a single one, we finally process the binary music predictions using a morphologic closing of 3 seconds.

\section{Implementation Details}
\label{sec:app_implementation}

In videos whose total duration is longer than $K\cdot t$, clips are sampled from a subset of the video, of duration $K\cdot t$. During training, this subset is sampled randomly, and at test time the subset is found deterministically by centering it in the middle of the video. 

The Transformer architecture follows the same design as the Transformer Encoder in the original paper \cite{Vaswani2017}. We use hidden dimensionality $d_h=256$, two layers, and two heads. We adapt the dimensionality $d_{\text{in}}$ of the input features to the hidden size $d_h$ using a linear projection layer.

We train the model using backpropagation, and following the optimization parameters in \cite{touvron2020deit}. Specifically, we train all the models using AdamW optimizer and a cosine learning rate annealing strategy with an initial value of 1e-3. We set the total batch size to 2048. We implement our models using PyTorch \cite{NEURIPS2019_9015}, and train on 8 RTX 2080Ti GPUs for 2 days.

\section{Other Experiments}
\label{sec:app_experiments}

\subsection{Experiment Setup Details}

In order to obtain audios of guitars (used in Figure 8 of the main paper), we use the MedleyDB dataset \cite{bittner2014medleydb}, which has separate music tracks for every instrument, for a series of 196 songs (including both version 1 and 2 of the dataset).

\subsection{Human Experiment Details}
\label{appx:human}

Each human was shown 20 example video pairs for each task (video-to-music, and music-to-video), where the pair consisted of the result provided by our model and the result provided by the baseline. The two clips in the pair contained either the same video and different music (for the video-to-music task), or two different videos with the same music (for the music-to-video task). The task consisted of a binary choice between the two clips, according to the best fit between music and video in the clip. Every human was shown different examples from the test set.

We collected a total of 296 human evaluations for the video-to-music task, with 220 of them (74.3\%) preferring our model's result over the baseline. Similarly, we collected 372 human responses for the music-to-video task, with 257 (69.1\%) preferring our model's output. Using the binomial test for statistical significance, we obtain in both cases a \emph{p}-value well below 0.01, validating the claim that our model is preferable to humans.

\subsection{Quantitative Analysis Extended}
\label{app:quantitative}
\definecolor{mycolor}{HTML}{143a61}
\definecolor{mycolor2}{HTML}{FF5733}
\definecolor{mycolor3}{HTML}{AD23AD}

\begin{figure}
\centering
\begin{tikzpicture}
    \pgfplotsset{%
        width=1.\columnwidth,
        height=0.6\columnwidth
    }
    \begin{axis}[
        legend style={at={(0.02,0.25)},anchor=west},
        xlabel=Ratio $r$ with respect to original value,
        ylabel=Recall@10,
        xmin=0.5, xmax=1.5,
        ymin=20, ymax=50,
        xtick={0.6, 0.8, 1, 1.2, 1.4},
        xticklabels={$\times$0.6, $\times$0.8, $\times$1.0, $\times$1.2, $\times$1.4},   % <---
        ytick={0,10,...,100},
        every axis plot/.append style={ultra thick}
                ]
    \addplot[smooth,mycolor] plot coordinates {
        (0.7, 36.955)
        (0.9, 40.933)
        (1, 42.37)
        (1.1, 41.522)
        (1.3, 38.905)
    };
    \addlegendentry{Tempo modification}
    
    \addplot[smooth,,mycolor2] plot coordinates {
        (0.6, 41.24)
        (0.8, 42.05)
        (1, 42.37)
        (1.1, 42.30)
        (1.3, 41.74)
    };
    \addlegendentry{Brightness modification}
    
    \addplot[smooth,,mycolor3] plot coordinates {
        (0.75, 37.74)
        (0.9, 39.32)
        (1, 42.37)
        (1.1, 39.53)
        (1.25, 36.84)
    };
    \addlegendentry{Hue modification}

    \end{axis}
\end{tikzpicture}
\caption{\textbf{Manipulation of tempo and color brightness}. Both attributes influence, but are not critical to, model performance.} 
\label{fig:tempo_brightness}
\end{figure}
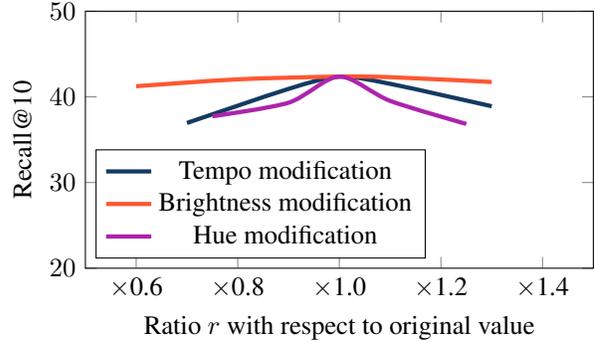

\begin{figure*}[t]
\includegraphics[width=0.8\textwidth]{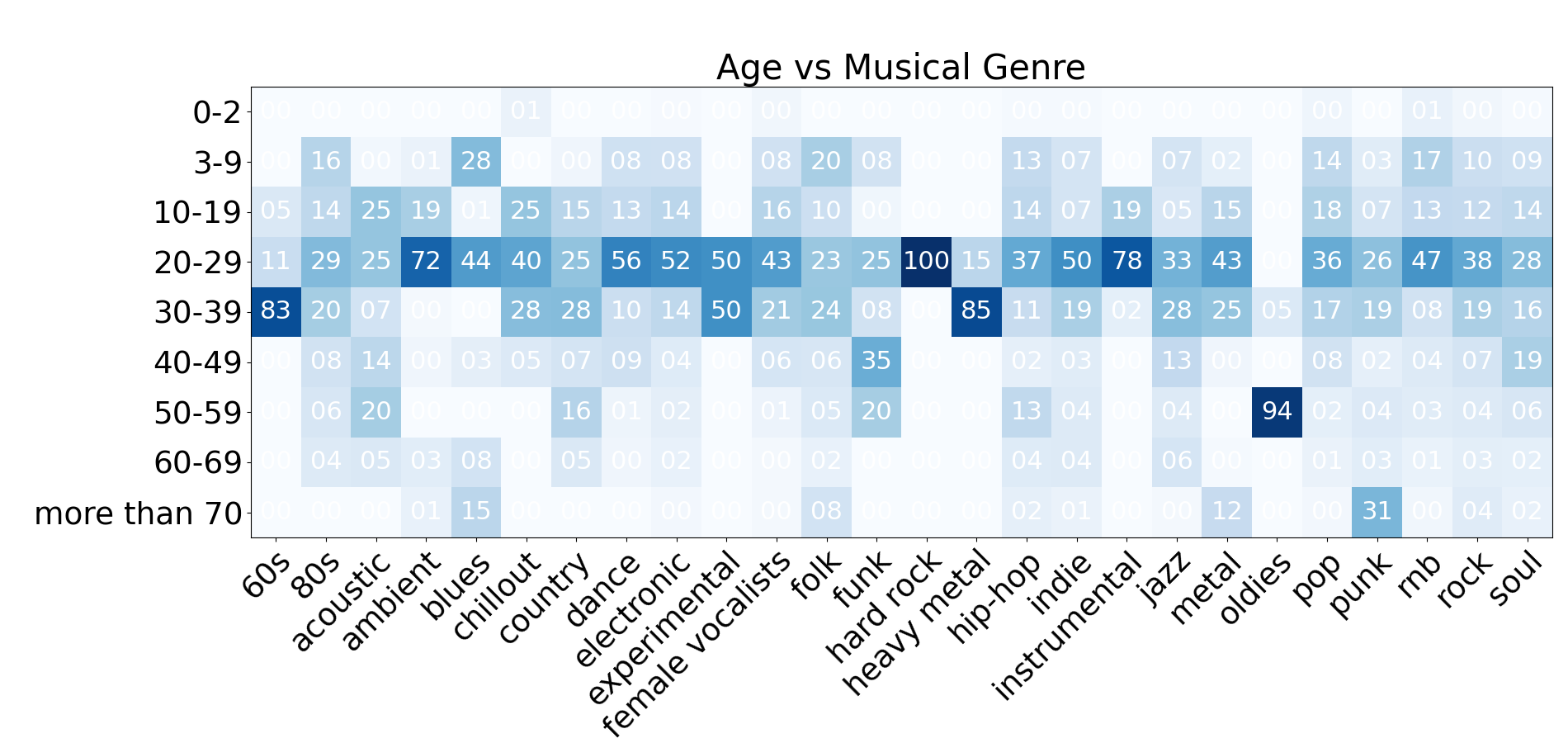}
\centering
\caption{Age vs genre. This figure is equivalent to Figure 4 in the main paper, but showing age targets, instead of gender or race.}
\label{fig:age}
\end{figure*}

\begin{figure*}[t]
\vspace{-2cm}
\includegraphics[width=0.8\textwidth]{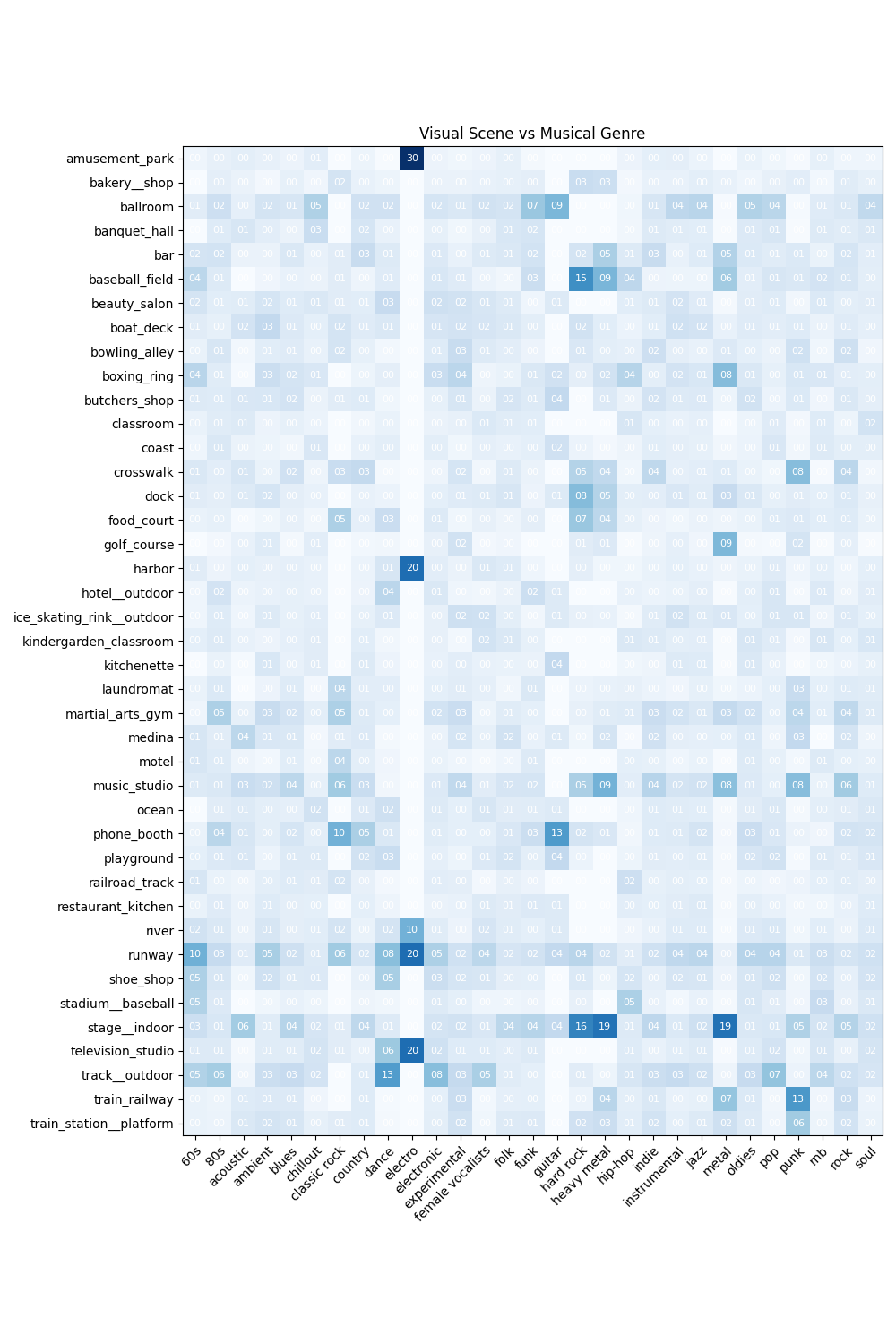}
\centering
\vspace{-2cm}
\caption{Scene vs genre. This figure is equivalent to Figure 4 in the main paper, but showing visual scene targets, instead of gender or race. While hard to associate specific genres to specific scenes, we note that there is a correlation, and the model is picking up on some signal regarding visual scene.}
\label{fig:scene}
\end{figure*}

\begin{figure*}[t]
\includegraphics[width=0.8\textwidth]{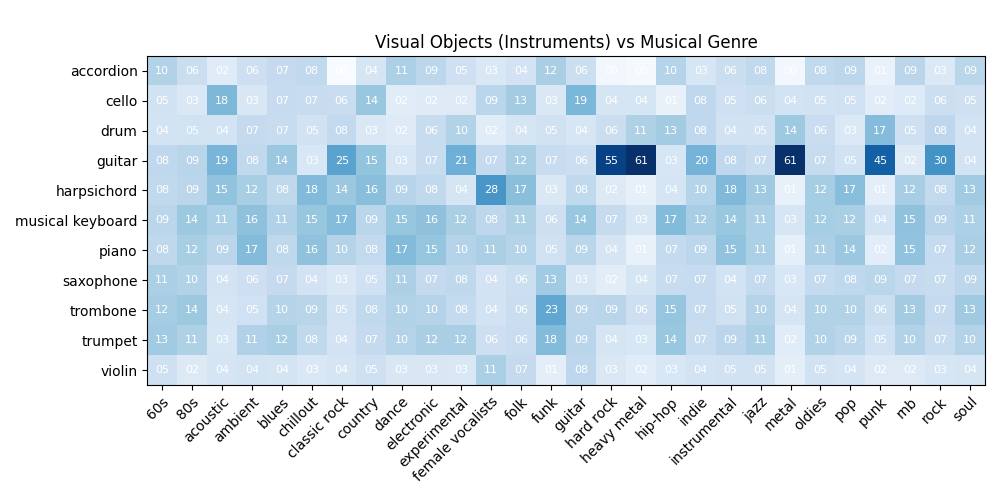}
\centering
\caption{Instrument vs genre. This figure is equivalent to Figure 4 in the main paper, but showing instrument targets, instead of gender or race. The most retrieved instrument is the guitar, suggesting that the model has a good representation of guitars, and therefore we can use guitars for our conditioning experiments.}
\label{fig:instrument}
\end{figure*}

In this section, we extend the analysis in Section~\ref{sec:analysis} in the main paper (``Quantitative Analysis''). 

First, we show the curves of accuracy versus change rate $r$ in color brightness, color hue, and music tempo in Figure~\ref{fig:tempo_brightness}. In the case of hue, which is not an intensity parameter, a change of $r$ is a change in the hue following the equation $h_\text{new}=(h_\text{original}+ 360(r-1))\mod360$, where $h$ represents hue, which is an angle with values in $[0,360)$. This hue change formulation allows us to represent all three modifications in a common scale.

We also show the matrices relating age, visual scene, and visual objects (instruments) to music genre in \Cref{fig:age,fig:scene,fig:instrument}. 
For the visual scene analysis on Places, from the 201 we only show the classes for which the sum of all the genres provides at least a sum of 20\%. To obtain the balanced dataset of musical instruments, we download the images in the Open Images Dataset \cite{kuznetsova2018open} that contain bounding box annotations of musical instruments and crop the instruments in those images. Then, we randomly sample 50 images from every instrument class in order to have a balanced dataset, %Instruments with less than 100 images in the dataset are not used in our analysis. 
and proceed as in the other attribute studies.

We also implemented a same-genre baseline, where we assume the genre label for each video and music track is known (for the video, we use the genre of its associated GT music). From the full set of target candidates, we take just the ones with the same genre as the query, and randomly select one as the prediction. The average R@10 values are 10.00\% for segment-level (compared to our 42.37\%), and 7.27\% for track-level (compared to our 42.27\%). This result shows that, while genre is an important attribute, our model captures a variety of other relevant audiovisual cues ({\em on top of} capturing genre information). Note that the model does not have access to the ground truth (GT) genre. Recognizing the genre is a hard task by itself, and only well defined for music (not for video).

\subsection{Qualitative comparison to baselines}
\label{appx:qualitative}

We perform some qualitative comparisons between the GT audios and videos, the suggested baseline, and our model results, on the examples shown in Figure 1 of the main paper. 
% The first video-to-music example uses this video as the query: YouTube ID \href{https://www.youtube.com/watch?v=4IUqQ5GT4y0&t=80s}{4IUqQ5GT4y0}, where the GT audio can be heard. 
In the first video-to-music example, the ground truth genre is ``rock''. Compared to the ground truth audio, the music retrieved by our method %(heard in the suppl.\ material) 
has a very similar tempo, an equal reliance on pronounced beats, and a very similar language (Spanish vs.\ Portuguese). %Interestingly, our retrieved music track is classified as Hip-Hop. 
In contrast, the same-genre baseline (a random retrieval of a rock track) returns a music track %with Youtube ID \href{https://www.youtube.com/watch?v=ExckSO5BFjg&t=52s}{ExckSO5BFjg} 
which, while having similar instrumentation to the ground truth (electric guitar and drums), is otherwise very different in terms of mood (sadder, less active), tempo (much slower), and language (Slovenian), and is a poor overall fit for the query video. We generally found that the same-genre baseline, unlike our model, is incapable of capturing important attributes not directly related to genre.

\subsection{t-SNE visualizations} 
\label{appx:tsne}

\begin{figure*}[t]
\includegraphics[width=0.8\textwidth]{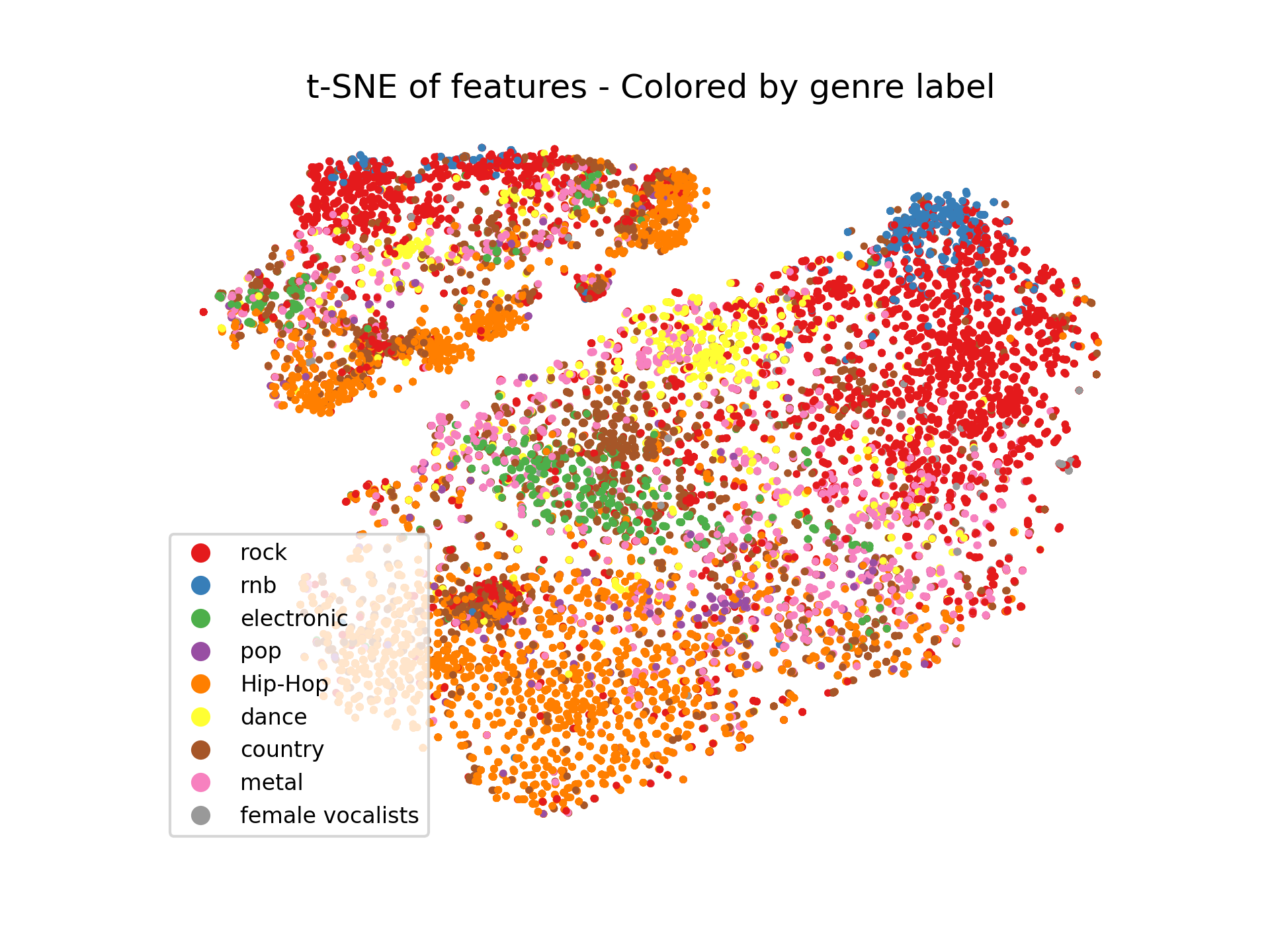}
\centering
\caption{We project the music features into a 2-dimensional space using t-SNE \cite{van2008visualizing}, and color the data points by genre label. Music tracks are generally clustered by genre.}
\label{fig:tsne}
\end{figure*}

We computed t-SNE visualizations \cite{van2008visualizing}, which show videos and music tracks are generally clustered by genre, although not exclusively. See Fig.~\ref{fig:tsne}.

\subsection{Addressing Potential Learning Shortcut}

\begin{table*}[t]
\small
\setlength\extrarowheight{-3pt}
\centering
\caption{\label{tab:shortcut}Track-level retrieval results for MusicVid-YT8M when exploiting the potential shortcut.}
\vspace{-0.2cm}
\begin{tabular}{lrr rrr rrr r}
\toprule
{}                                              & \multicolumn{2}{c}{Median Rank ${\downarrow}$} &   \multicolumn{7}{c}{Recall ${\uparrow}$} \\
\cmidrule(lr){2-3}
\cmidrule(lr){4-10}

{}                                              & V$\rightarrow$M & M$\rightarrow$V &  \multicolumn{3}{c}{ V$\rightarrow$M} & \multicolumn{3}{c}{M$\rightarrow$V} & Average \\
\cmidrule(lr){4-6}
\cmidrule(lr){7-9}
{}                                              &  &  &     R@1 &     R@5 &    R@10 &   R@1 &   R@5 &  R@10 & R@10 \\

\midrule
Baseline + CLIP and DeepSim features                       & 5 & 5 & 20.31 & 50.96 & 67.43 & 23.49 & 55.52 & 70.56 & 71.64 \\
MVPt (ours)                                               & \textbf{1} & \textbf{1} & 48.07 & 85.44 & 90.13 & 48.08 & 85.74 & 90.39 & \textbf{90.26} \\
\midrule
Chance                                                                 & 48 & 48 &	3.01 & 9.86 & 17.40 & 3.01 & 9.86 & 17.40 & 17.40 \\
\bottomrule
\end{tabular}
\end{table*}
% \vspace{-0.2cm}

Transformers take in a sequence of clips, instead of a single average of all clips. This makes them capable of obtaining information about the length of the sequence. Therefore, a potential learning shortcut Transformers may exploit is the use of sequence length as a signal to match music to a video (music associated to a specific video will have the same length than the video). In order to show that our Transformer model is not relying solely on this shortcut and it is mostly exploiting other cues, we run a test that controls for the sequence length. 
Namely, we constrain the candidate retrieval set to be the same length as the query by setting the similarity scores between different-length sequences to zero.
% Namely, we set all similarity scores between different-length sequences to zero, so that the pool of candidates to retrieve from is reduced to just those that have the same length as the query. 
This would be equivalent to detecting the shortcut and exploiting it with perfect accuracy. 

We perform this test for the YT8M-MusicVideos dataset, at the segment level (equivalently to Table 1 in the main paper). The numbers are in Table~\ref{tab:shortcut}. Overall, the numbers imply that 1) our model is not relying on this shortcut, and 2) even when controlling for the bias (and providing the MLP baseline with a tool that it cannot have by construction), the Transformer model is still significantly better than the MLP baseline, showing that modeling of temporal context in a non-trivial way is highly beneficial for the performance of the model.

% --- uncomment this to read the instructions
% \input{X_instructions}

\end{document}